\documentclass[useAMS,usenatbib]{mn2e}
\usepackage{txfonts}
\usepackage{graphicx}
\usepackage{verbatim}
\usepackage{rotating}
\usepackage[flushleft]{threeparttable}
\usepackage{color}

\newcommand\finetilde{{\raise.17ex\hbox{$\scriptstyle\sim$}}} 

\begin{document}

\newcommand{\qq}{\symbol{34}} 

\title[Profile Stochasticity in PSR J1909$-$3744]{Profile Stochasticity in PSR J1909$-$3744}
\author[L. Lentati et al.]{\parbox{\textwidth}{L. Lentati$^{1}$\thanks{E-mail:
ltl21@cam.ac.uk}, R. M. Shannon$^{2,3}$}\vspace{0.4cm}\\ %
$^{1}$ Astrophysics Group, Cavendish Laboratory, JJ Thomson Avenue,  Cambridge, CB3 0HE, UK\\
$^{2}$ CSIRO Astronomy and Space Science, Australia Telescope National Facility, Box 76 Epping, NSW, 1710, Australia\\
$^{3}$ International Centre for Radio Astronomy Research, Curtin University, Bentley, WA 6102, Australia}
\maketitle

\label{firstpage}

\begin{abstract}
We extend the recently introduced Bayesian framework `Generative Pulsar Timing Analysis' to incorporate both pulse jitter (high frequency variation in the arrival time of the pulse) and epoch to epoch stochasticity in the shape of the pulse profile.  This framework allows for a full timing analysis to be performed on the folded profile data, rather than the site arrival times as is typical in most timing studies.  We apply this extended framework both to simulations, and to an 11 yr,  10 cm data set for PSR J1909$-$3744.  Using simulations, we show that temporal profile variation can induce timing noise in the residuals that when performing a standard timing analysis is highly covariant with the signal expected from a gravitational wave (GW) background.  When working in the profile domain, these variations are de-correlated from the expected GW signal, resulting in significant improvement in the obtained upper limits. Using the PSR J1909$-$3744 data set from the Parkes Pulsar Timing Array project, we find significant evidence for systematic high-frequency profile variation resulting from non-Gaussian noise in the oldest observing system, but no evidence for either detectable pulse jitter, or low-frequency profile shape variation.  Using our profile domain framework we therefore obtain upper limits on a red noise process with a spectral index of $\gamma = 13/3$ of $1\times10^{-15}$, consistent with previously published limits.
\end{abstract}


\section{Introduction}

The emission of electromagnetic radiation from pulsars in the radio band is thought to be the result of highly energetic charged particles being accelerated, and then escaping, along open magnetic field lines in the pulsar's magnetosphere.  When the magnetic axis is not perfectly aligned with the rotational axis, the result is a beam of radiation that sweeps across the sky.  If an observer lies in the path of this beam, they will observe a series of regularly timed pulses, giving the pulsar its trademark lighthouse effect.

The time of arrival (TOA) of these pulses is predicted by the pulsar's `timing model'.  This is a deterministic description of the pulsar, describing its rotational period and spin down, astrometric properties, and if the pulsar is part of a binary system, additional parameters such as its orbital period, and geometry of the orbit with respect to the Earth.

When a pulsar is observed for the purpose of timing, the individual pulses from a single observing epoch are folded together using some fiducial timing model to form an average pulse profile for that observation.  This averaged data can then be compared to a model for the pulse profile, a process that results in the creation of a set of observed TOAs, referred to as  site arrival times (SATs).   These SATs can then be compared to those predicted by the pulsar's timing model, where the difference between the two are known as the `timing residuals'.  These residuals carry physical information about the unmodelled effects in the pulse propagation, including intrinsic variation in the pulse emission time  \citep[e.g.][]{2014MNRAS.443.1463S,2010MNRAS.402.1027H}, or extrinsic factors, such as perturbations due to gravitational waves (GWs) \citep[e.g.][]{1978SvA....22...36S,1979ApJ...234.1100D}.

\begin{figure}
\begin{center}$
\begin{array}{c}
\hspace{-1cm}
\includegraphics[width=100mm]{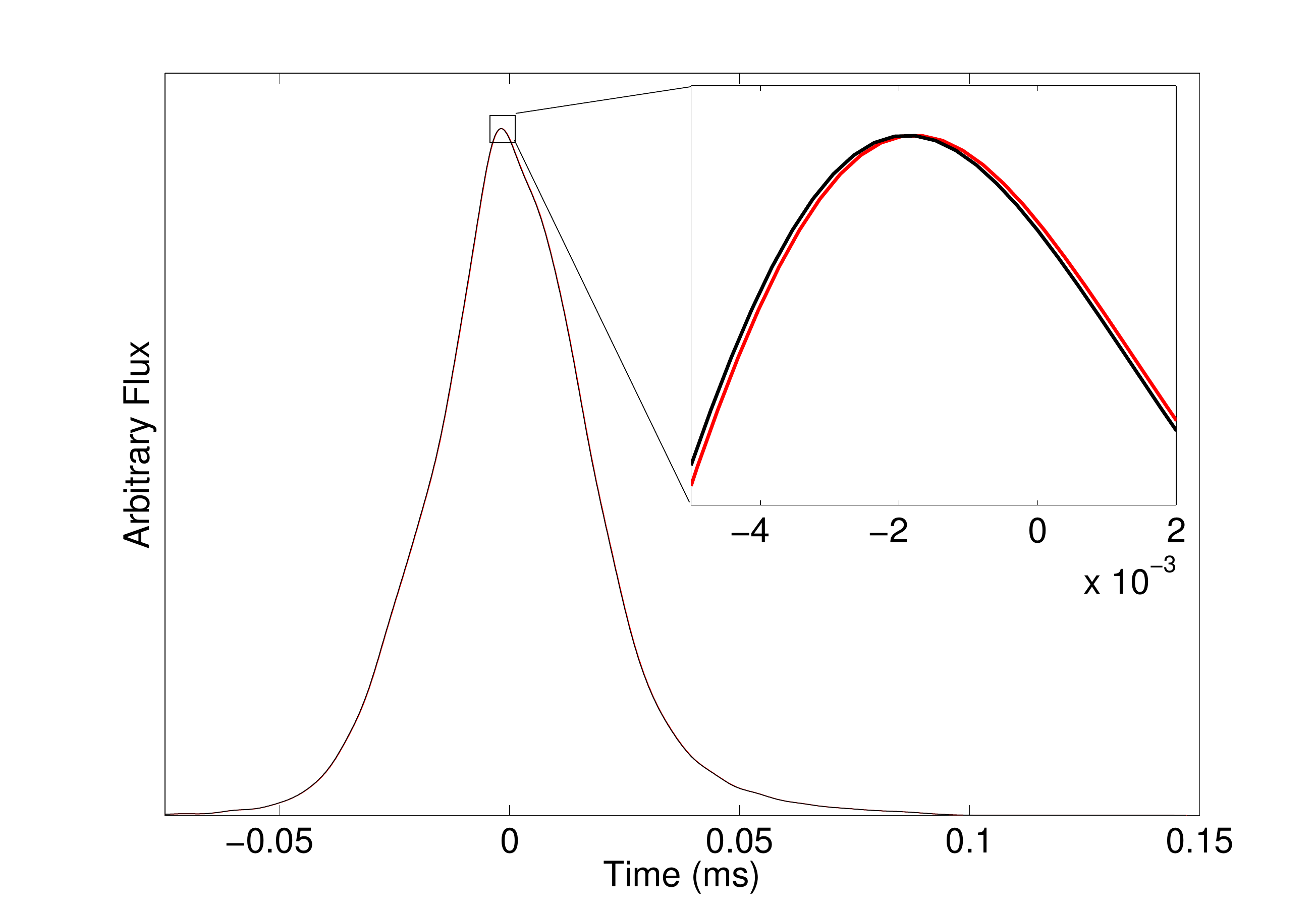} \\
\end{array}$
\end{center}
\caption{Noiseless model for the deterministic profile of PSR J1909$-$3744 at zero phase (black line), and after being shifted by 100ns, comparable to the effect of a passing GW from a $10^{-15}$ isotropic GWB (red line).  The timespan along the x-axis covers ~7.5\% of a full rotation for this pulsar.  In the PSR J1909$-$3744 10cm data set described in Section \ref{Section:Dataset} this corresponds to, at best, $\sim$ one tenth of a phase bin.}
\label{Fig:GWProfile}
\end{figure}

The most stringent $95\%$ upper limit on the amplitude of an isotropic stochastic GW background (GWB) formed from the incoherent superposition of GWs emitted from a large number of merging supermassive black-hole binaries is $A < 1\times10^{-15}$ for a reference frequency of one~yr$^{-1}$ (\citealt{2015Sci...Shannon}, henceforth S15). This is equivalent to a red noise process with an rms amplitude of $\sim$ 70~ns in a 10 yr data set.  In Fig. \ref{Fig:GWProfile}, we show the effect of a 100ns deviation due to a GWB on the arrival time of the average profile at a single observational epoch.  In the PSR J1909$-$3744 10cm data set described in Section \ref{Section:Dataset} the largest number of phase bins used per epoch is 2048, corresponding to a deviation of only $\sim$ one tenth of a phase bin.

Several approaches to building a profile template have previously been used.  For example, a template can be formed by averaging over the individual profiles at each observational epoch to create a single high S/N profile that can then be used to form the TOAs directly. This, however, has the disadvantage that any noise present in the template can bias the resulting TOA estimates \citep{2005MNRAS.362.1267H}, especially if the template noise is correlated with noise in the profiles. Alternatively, rather than use the averaged data,  a smoothed version of the template can be used \citep[e.g.][]{2013ApJ...762...94D}, or analytic functions can be fit to the averaged profile to form a noise free template \citep[e.g.][]{2013PASA...30...17M}.

\begin{figure*}
\begin{center}$
\begin{array}{c}
\hspace{-1cm}
\includegraphics[width=150mm]{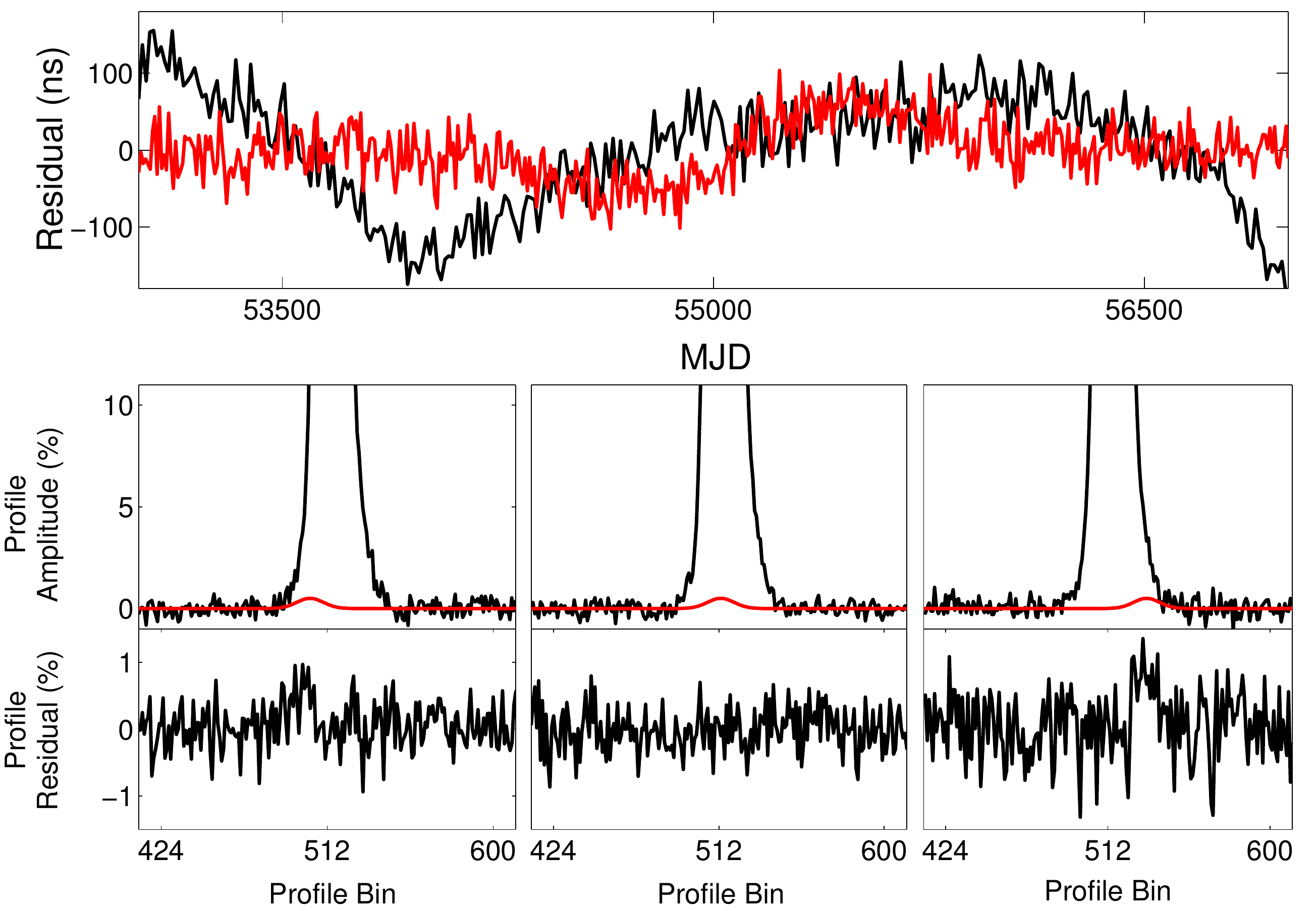} \\
\end{array}$
\end{center}
\caption{(Top) Black line - Simulated residuals due to a GW signal from an isotropic stochastic background with an amplitude of $1\times10^{-15}$, consistent with the most stringent 95\% upper limits set by Shannon et al. (2015).  TOAs were simulated using the highest signal to noise profile in the PSR J1909$-$3744 data set used in Section \ref{Section:Dataset}, resulting in uncertainties of 20ns for each observation. Red line - Simulated residuals induced by the passage of an additional Gaussian component to the profile data (see bottom panels), not included in the template at the time of forming the TOA, with an amplitude of 0.5\% that of the observed profile.  The two signals are of comparable amplitude, implying that any unmodelled profile variation larger than this will quickly dominate over a GW signal in the TOAs.   (Bottom)  3 examples of the additional Gaussian component at different positions in the main profile, and residuals from the profile fit.}
\label{Fig:MoneyPlot}
\end{figure*}

Once a template has been developed it is then used to form the TOAs for each observational epoch.  This is most commonly done via the `Fourier phase-gradient method' \citep{1992RSPTA.341..117T} in which the phase offset between the two is computed using the Fourier transform of both the template, and the profile at each epoch, and a cross correlation between the two performed.  Alternative time domain approaches have also been used \citep[e.g.][]{2005MNRAS.362.1267H}, however regardless of the approach,   they all share a common assumption; that the profile is stable within radiometer noise from epoch to epoch.

While long term stability of pulse profiles has been shown in some pulsars (e.g. \citealt{2013CQGra..30p5019S}), epoch to epoch variation in the profile shape has also been observed.  For example, a study of morphological variability in PSR J1022+1001 suggests that the pulse profile varies at the few per cent level \citep{2004MNRAS.355..941H, 2015MNRAS.449.1158L}, while PSR J0437$-$4715 has also been observed to show timing instability \citep{2006MNRAS.369.1502H}.  In both cases the origin of the instability could be instrumental, for example, due to polarization calibration errors, or it could be the results of the intrinsic stochasticity of the profile. The individual pulses from a pulsar are known to show a high degree of variability (e.g. \citealt{1981ApJ...249..241H}), and so as instrumentation improves and radiometer noise decreases, this intrinsic stochasticity will unavoidably become more significant within a single observation. Profile variability has also been observed in young pulsars, where in some cases timing noise has been found to be correlated with changes in the pulse shape \citep{2010Sci...329..408L}.  Pulse profile variability associated with instrumental distortions has also been widely observed, particularly in jitter-dominated observations of young pulsars or with instruments with low-bit digitisation \citep{1998PASP..110.1467J}.   Typically the effects of these distortions have been modelled in the TOA domain.  This is done both by including additional white noise parameters referred to as `EFAC' and `EQUAD', which scale and add in quadrature to the formal TOA uncertainties, and by incorporating a model for low frequency timing noise into the analysis (e.g. \citealt{2011MNRAS.418..561C,2013MNRAS.428.1147V, 2014MNRAS.437.3004L}).  This has the disadvantage that in a single pulsar these distortions could be covariant with the GW signal (see Figure \ref{Fig:MoneyPlot}).

In the top panel of Fig. \ref{Fig:MoneyPlot}, we compare simulated residuals induced by the GW signal from an isotropic stochastic background with an amplitude of $1\times10^{-15}$ (black line),  with those that result from the passage of an additional Gaussian component through the profile, with an amplitude of 0.5\% that of the observed profile, which is not appropriately modelled by the single average profile  used to form the TOAs (see bottom panels).  All TOAs were simulated using the highest signal to noise profile in the PSR J1909$-$3744 data set used in Section \ref{Section:Dataset}, resulting in uncertainties of 20ns for each observation. The two signals are of comparable amplitude, implying that any unmodelled profile variation larger than this will quickly dominate over a GW signal in the TOAs.   

In \cite{2015MNRAS.447.2159L} (henceforth L15) a Bayesian framework was introduced dubbed `Generative Pulsar Timing Analysis' (GPTA) that allows for a full timing analysis using the folded profile data, rather than the SATs that result from the cross correlation with a profile template.  This allowed for analysis of the pulsar's timing model, along with intrinsic stochastic processes such as spin noise -- low frequency variation in the pulse TOAs -- simultaneously with a model for the pulse profile, for which a shapelet basis was used.

In this work we extend this framework to incorporate epoch to epoch changes in the profile.  We include a model for pulse jitter -- high frequency stochastic variation in the arrival time of the profile model -- along with models for variations in the shape of the profile, which we obtain by calculating the power spectrum of the variance in our shapelet model as a function of scale in phase space.  While this doesn't constitute a physical model for the epoch to epoch stochasticity, by obtaining the power spectrum of the variations, we can begin to characterise the shape changes in a statistically robust manner, ultimately leading to a better understanding of their origins. 

In Sections \ref{Section:Like} to \ref{Bayesian} we describe the models used in our profile domain analysis, and how we implement them in our Bayesian framework.  In Section \ref{Section:Dataset} we describe the 10cm PSR J1909$-$3744 data set that we use to construct our simulations described in Section \ref{Section:Simulations}, and that we use in our analysis in Section \ref{Section:RealData}.  Finally we offer some concluding remarks in Section \ref{Section:Conclusions}.

\section{A profile domain model}
\label{Section:Like}

The methods used in this analysis are drawn from those presented in L15.  Here, our pulsar timing analysis is performed entirely with the profile data, rather than the TOAs formed from those profiles.  Qualitatively, in each likelihood calculation, we construct a model for the deterministic (or average) profile using a shapelet basis, and generate a model time of arrival at each observational epoch for that profile using the pulsars timing model.  Both these steps occur simultaneously, such that both the parameters that describe the shapelet model, and the timing model parameters are free to vary within our analysis.  

While a full description of the general framework we will use is available in L15, in this work we will be extending the methodology significantly to incorporate the possibility of epoch to epoch variation in the profile.  We include models for pulse jitter -- a shift in the arrival time of the deterministic profile -- as well as shape variation that could be of instrumental, or astrophysical origin.  As such we will include below an overview of the basic framework, before providing details on the modifications required to incorporate profile stochasticity.  

\subsection{Shapelets}
\label{section:Shapelets}

A thorough description of the Shapelet formalism can be found in \cite{2003MNRAS.338...35R}, with astronomical uses being described in e.g, \cite{2004AJ....127..625K,2013MNRAS.430.2454L,2003MNRAS.338...48R}.  Here we give only an outline to aid later discussion.

Shapelets are described by a set of dimensionless basis functions, which in one dimension can be written as:

\begin{equation}
\phi_n(x) \equiv \left[2^nn!\sqrt{2\pi}\;\right]^{-1/2} H_n\left(\frac{x}{\sqrt{2}}\right)\;\exp\left(-\frac{x^2}{2}\right),
\end{equation}
where $n$ is a non-negative integer, and $H_n$ is the Hermite polynomial of order $n$.  Therefore the lowest order shapelet is given by a standard Gaussian ($H_0(x) = 1$), with higher order terms represented by a Gaussian multiplied by the relevant polynomial.

These are then modified by a scale factor $\Lambda$ which is a free parameter to be fitted for, in order to construct dimensional basis functions:

\begin{equation}
B_n(x;\Lambda) \equiv \Lambda^{-1/2} \phi_n(\Lambda^{-1}x).
\end{equation}
These basis functions are orthonormal, i.e:

\begin{equation}
\int_{-\infty}^\infty \; \mathrm{d}x \; B_n(x;\Lambda)B_m(x;\Lambda) = \delta_{nm},
\end{equation}
where $\delta_{nm}$ is the Kronecker delta, so that we can represent a function $s(x)$ as the sum:

\begin{equation}
\label{Eq:oldshapefunction}
s(x, \mathbf{\zeta}, \Lambda) = \sum_{n\mathrm{=0}}^{n_{\mathrm{max}}} \zeta_nB_n(x;\Lambda),
\end{equation}
where $\zeta_n$ are the shapelet amplitudes, and $n_{\mathrm{max}}$ the number of shapelet basis vectors included in the model.  

In our analysis of pulsar timing data, we form a single profile shape, which will then be scaled from epoch to epoch.  We therefore redefine Eq. \ref{Eq:oldshapefunction}, such that we have a single global amplitude $A$, and $n_{\mathrm{max}}-1$ parameters $\zeta_n$ which are the amplitudes for the shapelet components that have $n>0$.  These therefore represent the relative contribution to the overall profile shape, relative to the zeroth-order term, which we take to have an amplitude of 1.  Written in this way Eq. \ref{Eq:oldshapefunction} becomes:

\begin{equation}
\label{Eq:shapefunction}
s(x, A, \mathbf{\zeta}, \Lambda) = A\sum_{n\mathrm{=0}}^{n_{\mathrm{max}}} \zeta_nB_n(x;\Lambda).
\end{equation}

Finally, the total integrated flux in the model profile, $F_{\mathrm{tot}}$, is given by

\begin{equation}
F_{\mathrm{tot}} =\;\int_{-\infty}^{\infty} \;\mathrm{d}x \; s(x) \; = \; A\sum_{n\mathrm{=even}} \; \zeta_n\big[2^{1-n}\sqrt{2\pi}\Lambda\big]^{\frac{1}{2}} {n \choose n/2}^{\frac{1}{2}}.
\end{equation}

While there is no natural basis in which to describe the shape of the pulse profile, shapelets present several advantages when compared to, for example, a set of von Mises functions.  A shapelet model requires a single width parameter, and a set of $n_\mathrm{max}$ amplitudes which multiply the linear basis vectors given that width.  Phase wrapped Gaussians, or von Mises functions, however, have $n$ widths and $n-1$ relative positions, which can become highly multi-modal (each component used can switch places while giving the same result), and thus computationally expensive when exploring the parameter space. In addition, as we will show in Section \ref{Section:Stochastic} it is trivial to use shapelets to incorporate additional stochastic elements in the profile, without having to sample numerically over a large number of additional dimensions, as we can deal with many of the amplitude parameters analytically.

Finally we note that while in this work we discuss only the use of one-dimensional shapelets, in \cite{2003MNRAS.338...35R} a two-dimensional shapelet basis is also described.  In the context of profile domain pulsar timing analysis, the use of two-dimensional shapelets can readily be seen as an extension to the methodology we will describe below.  While we will consider only radio data that has been fully integrated over frequency, modern observing systems observe wide bandwidths. Two dimensional template fitting that incorporates a model for both the frequency dependence of the profile across the band (e.g \cite{2014ApJ...790...93P, 2014MNRAS.443.3752L}), and the broadband component of pulse jitter, could further improve the timing precision obtainable via a profile domain analysis. We will investigate this possibility in future work.

\subsection{Evaluating the timing model}

We begin by considering our data, $\mathbf{d}$, a set of $N_d$ integrated pulse profiles, where the profile at observational epoch $i$ consists of a set of $N_i$ values representing the flux density of the profile as measured at a set of times $\mathbf{t_i}$.  We represent $\mathbf{d}$ as a sum of both a deterministic and a stochastic component:

\begin{equation}
\mathbf{d} = \mathbf{s} + \mathbf{n},
\end{equation}
where $\mathbf{d}$ represents the $\sum_{i=1}^{N_d} N_i$ total data points for the $N_d$ pulse profiles for a single pulsar, with $\mathbf{s}$ and $\mathbf{n}$ the deterministic and stochastic contributions to the total respectively, where any contributions to the latter will be modelled as random Gaussian processes.

We first consider the stochastic component of the signal, $\mathbf{n}$, to be described solely by an uncorrelated random Gaussian process with RMS $\sigma_i$, determined individually at each epoch $i$. The deterministic component, $\mathbf{s}$,  consists only of (i) our shapelet model for the pulse profiles with the centroid position for each model profile determined using the set of arrival times $\mathbf{\tau}(\mathbf{\epsilon})$  predicted by the pulsar's $m$ timing model parameters, $\mathbf{\epsilon}$, and (ii) an arbitrary baseline offset for each profile epoch.  With the inclusion of the timing model, we can rewrite Eq. \ref{Eq:shapefunction} to describe the shapelet model for a particular epoch $i$ as:

\begin{equation}
\label{Eq:newshapefunction}
s_i(t, A, \mathbf{\zeta}, \Lambda, \epsilon) = A\sum_{n\mathrm{=0}}^{n_{\mathrm{max}}} \zeta_nB_n(t-\tau_i(\epsilon);\Lambda),
\end{equation}
with $\tau_i$ the barycentric arrival time for pulse $i$ predicted by the set of timing model parameters $\epsilon$.  

In L15 it was assumed that the correction to the SSB for each profile could be computed at the SAT obtained previously using traditional techniques for that profile.  In our analysis of the PSR J1909$-$3744, the inclusion of phase offsets between different observing systems (referred to as `Jumps') means that we must first correct the SATs using the values for these offsets, and the overall phase offset.  We then compute the correction to the Solar System Barycenter at these new values.  This is simply because the correction due to the Roemer delay can vary by $\sim 100$~ns across the set of times $\mathbf{t_i}$ that span the pulse period of the pulsar, which is significant in the context of the highest precision observations available today.

We can then write the likelihood that the data is described only by the shapelet parameters, $\mathbf{\theta} \equiv (A, \mathbf{\zeta}, \Lambda)$, the timing model parameters $\mathbf{\epsilon}$, and the baseline offset,  $\mathbf{\gamma}_i$, for each epoch $i$ as:

\begin{eqnarray}
\label{Eq:TimeLike}
\mathrm{Pr}(\mathbf{d} |\mathbf{\theta}, \mathbf{\epsilon}) &\propto& \prod_{i=1}^{N_d} \frac{1}{\sqrt{\mathrm{det}\mathbf{N_i}}}\\
&\times& \exp{\left[-\frac{1}{2}(\mathbf{d} - \mathbf{s_i}(\mathbf{\theta}, \mathbf{\epsilon}) - \mathbf{\gamma}_i)^T\mathbfss{N}_\mathbf{i}^{-1}(\mathbf{d} - \mathbf{s_i}(\mathbf{\theta},\mathbf{\epsilon}) - \mathbf{\gamma}_i)\right]} \nonumber,
\end{eqnarray}
where $\mathbfss{N}_\mathbf{i}$  is the white noise covariance matrix for epoch $i$, with elements $(N_i)_{jk} = \sigma_i\delta_{jk}$. 
%
%
%

\section{Including additional stochastic parameters} 
\label{Section:Stochastic}

\begin{figure}
\begin{center}$
\begin{tabular}{c}
\hspace{-3.5cm}
\includegraphics[width=80mm]{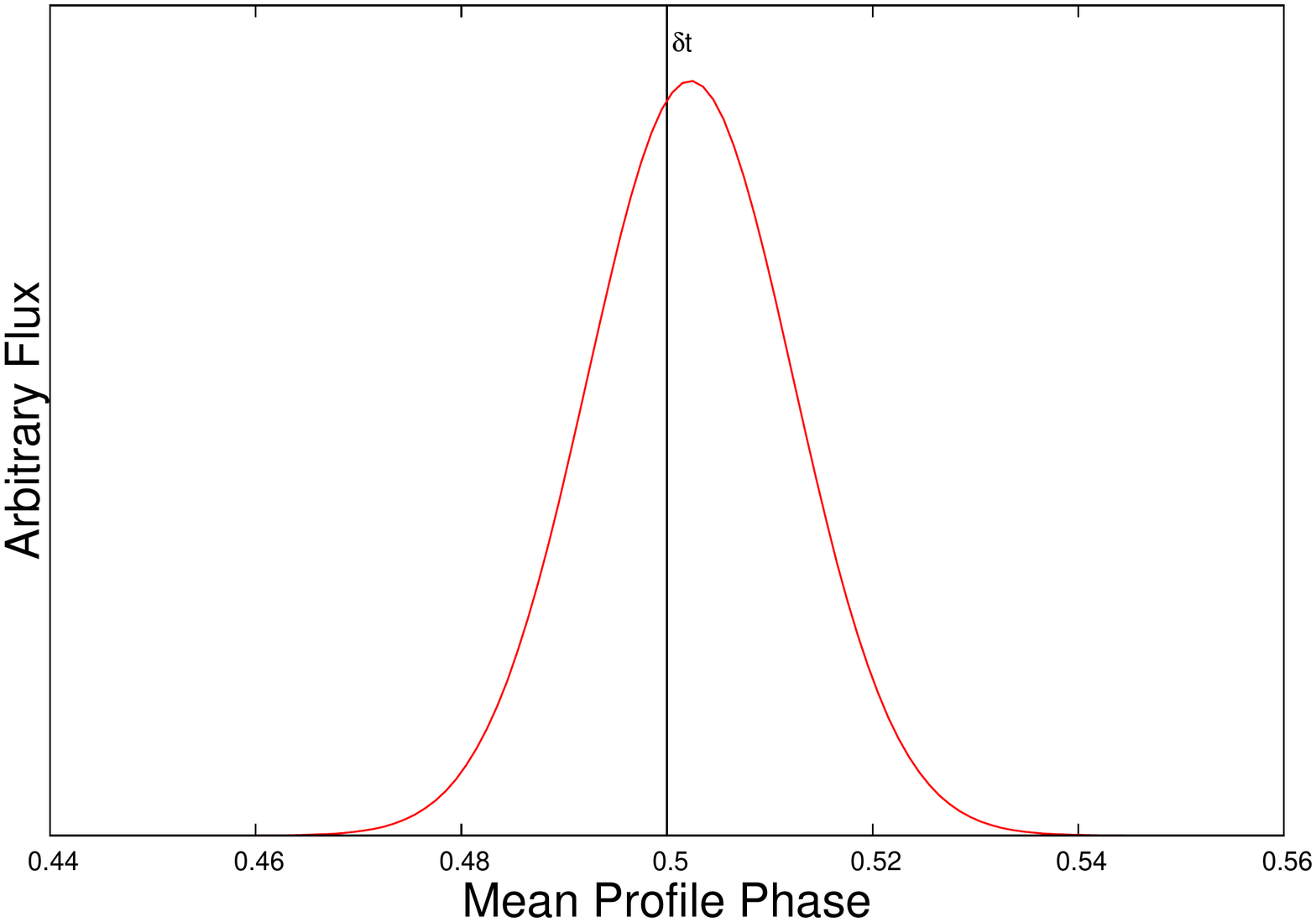} \\
\vspace{-1cm}
\hspace{-3.5cm}
\includegraphics[trim=50 50 50 100,clip,width=140mm]{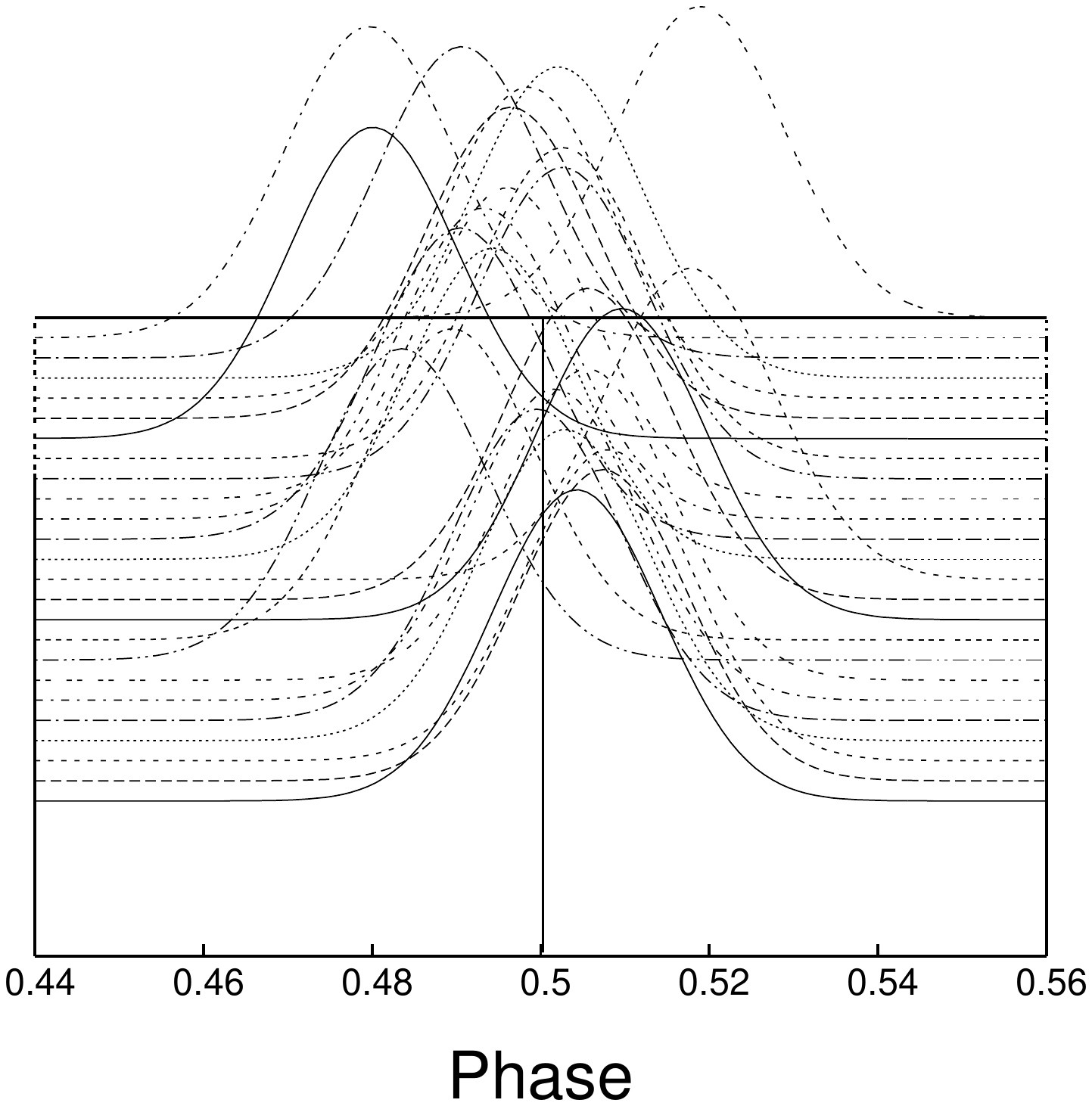} \\
\end{tabular}$
\end{center}
\caption{(Bottom) A series of simulated single pulses with a Gaussian profile, and a high frequency stochastic element to the pulse emission time.  (Top) The mean profile that results from averaging the single pulses.  Even if we assume that the mean profile is the same at each epoch, the observed arrival time could still be subject to a shift due to this stochasticity in emission.  In the TOA domain this is typically modelled with an EQUAD parameter, which adds in quadrature to the formal TOA uncertainty.  In the profile domain we can directly incorporate a model that describes a stochastic shift in the mean profile.}
\label{Fig:SinglePulsesPlot}
\end{figure}

\subsection{Radiometer noise}
\label{Section:Radiometernoise}

In Eq.~(\ref{Eq:TimeLike}) the amplitude of the radiometer noise in the profile data is taken to be a known quantity at each epoch.  In principle we would want to include $\sigma_i$ as a free parameter for each epoch, however this is not currently computationally tractable for large data sets.  In our analysis we can obtain a good initial estimate of the radiometer noise from the off-model region of the profile data, where the amplitude of the model profile is less than $0.1\%$.  As in TOA domain analysis we can then include additional parameters that help account for our uncertainty in this estimate, known as EFAC, which we denote $\alpha$, which scales our initial estimate of the radiometer noise for all epochs associated with a given observing system.

Our value for the radiometer noise then becomes:

\begin{equation}
\hat{\sigma}_{ij}^2 = (\alpha_i\sigma_{ij})^2.
\end{equation}

\subsection{Profile Jitter}
\label{Section:Jitter}

We define profile jitter in our model as a shift in the arrival time relative to those predicted by the pulsar's timing model of the deterministic profile that is uncorrelated between different observational epochs.  This could be the result of systematic effects, or intrinsic high frequency stochasticity  in the emission time of the pulse (see Fig. \ref{Fig:SinglePulsesPlot}, e.g. Shannon et al. (2014)).  In the latter case, even if we assume that the mean profile is the same at each epoch, the observed arrival time could still be subject to a shift due to this stochasticity in emission (see Fig. \ref{Fig:SinglePulsesPlot}).  Regardless of its origin, in the TOA domain this is typically modelled with an EQUAD parameter, which adds in quadrature to the formal TOA uncertainty.  In the profile domain we can directly incorporate a model that describes a stochastic shift in the mean profile.  In our model we assume that the jitter amplitudes at each epoch can be described by a Gaussian distribution, where the variance of that distribution is a free parameter in our analysis.

If we write the amplitude of the shift due to profile jitter at any epoch $i$ as $\delta t_i$, we can modify Eq. \ref{Eq:newshapefunction} trivially to include this new parameter as:

\begin{equation}
\label{Eq:jittershapefunction}
s_i(t, A, \mathbf{\zeta}, \Lambda, \epsilon, \delta t_i) = A\sum_{n\mathrm{=0}}^{n_{\mathrm{max}}} \zeta_nB_n(t-\tau_i(\epsilon)-\delta t_i;\Lambda).
\end{equation}
If this was the only modification to our analysis method, clearly introducing a shift parameter per epoch would result in complete covariance with the timing model parameters, it would be equivalent to introducing a jump in between each TOA.  We will later include a prior on these shift amplitudes that constrains them to come from a Gaussian distribution, and will show in the context of simulations, that when the jitter model is appropriate, this method produces results that are completely consistent with the simulated jitter amplitudes, and constraints that are superior to those obtainable in the TOA domain, with no loss in timing precision.

Written as in Eq. \ref{Eq:jittershapefunction} we have to include $N_d$ free parameters describing the jitter in our analysis, which, as for the overall profile amplitudes would rapidly become intractable for large data sets.  An advantage of the shapelet basis, is that for any arbitrary shapelet model we can analytically expand Eq. \ref{Eq:jittershapefunction} for small amplitudes $\delta t_i$ relative to the width parameter $\Lambda$.  This allows us to linearise the shift, and so marginalise analytically over the individual amplitudes, a process we will describe in Section \ref{section:Margins}.  A complete derivation of this linear translation operator, along with operators for other transformations, can be found in \cite{2003MNRAS.338...35R}.  The result is that, for any profile model $s_i(t, \mathbf{\zeta}, \Lambda, \epsilon)$, we can write down a `jitter profile' as:

\begin{eqnarray}
\label{Eq:JitterProfile}
j_i(t, \delta t_i, \mathbf{\zeta}, \Lambda, \epsilon) &=&  \frac{\delta t_i}{\sqrt{2}\Lambda}\sum_{n\mathrm{=0}}^{n_{\mathrm{max}}}\zeta_n\left(\sqrt{n}B_n(t-\tau_i(\epsilon);\Lambda)\right.\\
&-& \left.\sqrt{n+1}B_{n+1}(t-\tau_i(\epsilon);\Lambda)\right).
\end{eqnarray}

\begin{figure}
\begin{center}$
\begin{array}{c}
\includegraphics[width=90mm]{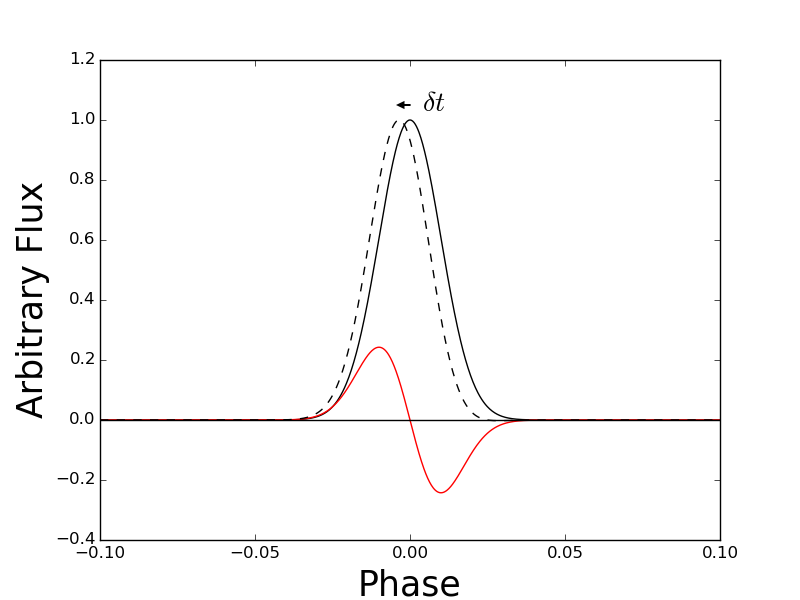} \\
\end{array}$
\end{center}
\caption{Example of the linear translation operator for the lowest order shapelet coefficient.  The black solid curve represents the model profile at the arrival time predicted by the pulsar's timing model.  The red line is the jitter profile defined in Eq.~(\ref{Eq:JitterProfile}).  Adding the jitter profile to the deterministic profile results in a shift by an amount $\delta t$ (black dotted line).  We stress that in addition to the shift amplitudes representing jitter, we enforce a strong additional constraint in the form of our prior in Eq.~(\ref{Eq:jitterPrior}).  All jitter amplitudes are constrained to come from a Gaussian distribution, whose variance is free parameter in our analysis.  We will show using simulations in Section \ref{Section:Simulations}, that when the stochastic component of the data really is well described by a shift in the arrival time, this profile domain model gives completely consistent constraints to those obtained in the TOA domain using an EQUAD parameter.}
\label{Fig:JitterExample}
\end{figure}

For example, in the case where $n_{max} = 0$, such that we only include the Gaussian term in our model, our jitter profile will be given by:

\begin{eqnarray}
j_i(t, \delta t_i, \mathbf{\zeta}, \Lambda, \epsilon) &=&  \frac{\delta t_i}{\sqrt{2}\Lambda}\zeta_0\left(-B_1(t-\tau_i(\epsilon);\Lambda)\right) \\
&=& -\frac{\delta t_i}{2\Lambda}\zeta_0(t-\tau_i(\epsilon))\exp\left(-\frac{1}{2}\frac{(t-\tau_i(\epsilon))^2}{\Lambda^2}\right)
\end{eqnarray}
which is equivalent to expanding $\exp\left(-\frac{1}{2}\frac{(t-\tau_i(\epsilon)-\delta t_i)^2}{\Lambda^2}\right)$ for small $\delta t_i$.  We illustrate this example in Fig. \ref{Fig:JitterExample}.

Our shifted shapelet model will then be given by:

\begin{equation}
\label{Eq:ShiftedShapelet}
s'_i(t, A, \delta t_i, \mathbf{\zeta}, \Lambda, \epsilon) = s_i(t, A, \mathbf{\zeta}, \Lambda, \epsilon) + j_i(t, \delta t_i, \mathbf{\zeta}, \Lambda, \epsilon).
\end{equation}

As mentioned previously, we must include a constraint on the amplitudes of the shift parameters $\mathbf{\delta t}$.  We do this by defining a prior on the amplitude parameters that is equivalent to fitting for the variance of the distribution.  This variance can be defined across an arbitrary set of observational epochs, for example, to model the variance separately as a function of observing system, or for all epochs simultaneously.

The covariance matrix of the jitter amplitudes, which we denote $\mathbfss{J}$,  is written:

\begin{equation}
\label{Eq:jitterPrior}
J_{ij} = \left< \delta t_i\delta t_j\right> = \mathcal{J}_i\delta_{ij},
\end{equation}
where the set of coefficients $\mathbf{\mathcal{J}}$ represent the theoretical variance of the jitter model at epoch $i$.

We normalise the amplitudes $\delta t_i$ at each epoch such that they can be described by a single variance $\mathcal{J}_i$ for all observations in units of seconds.  We cannot simply use the maximum likelihood amplitude of the profile model however, as if we believe our data to be affected by jitter, this amplitude will not be correct, which will decrease our sensitivity to the amplitude of jitter in the data set.

We therefore define the scaling factor $C_i$ such that:

\begin{equation}
\label{Eq:normalisation}
C_i = \frac{F_{\mathrm{d,}i}}{F_{\mathrm{tot,}i}},
\end{equation}
where $F_{\mathrm{tot,}i}$ is the total flux in the unscaled model profile at epoch $i$, and so is independent of the overall model amplitude, and $F_{\mathrm{d,}i}$ is an estimate of the total signal flux in the data at epoch $i$, which we calculate by integrating over the bins where the model profile is greater than 0.1\% of its total amplitude.

We can then write our joint probability density for the individual jitter amplitudes, and the variance parameters as:

\begin{eqnarray}
\label{Eq:Pulsarjitterlike}
\mathrm{Pr}(\mathbf{d} | \mathbf{\theta}, \bmath{\epsilon}, \mathbf{\delta t},  \mathbf{\mathcal{J}}) &=& \mathrm{Pr}(\mathbf{d} |\mathbf{\theta}, \bmath{\epsilon}, \mathbf{\delta t}) \times \mathrm{Pr}(\mathbf{\delta t} | \mathbf{\mathcal{J}})
\end{eqnarray}
where $\mathrm{Pr}(\mathbf{d} |\mathbf{\theta}, \bmath{\epsilon}, \mathbf{\delta t})$ has the same functional form as in Eq. \ref{Eq:TimeLike}, but with the the shapelet profile model being replaced with our shifted model in Eq.~(\ref{Eq:ShiftedShapelet}), and $\mathrm{Pr}(\mathbf{\delta t} | \mathbf{\mathcal{J}})$ is given by:

\begin{equation}
\mathrm{Pr}(\mathbf{\delta t} | \mathbf{\mathcal{J}}) \propto \frac{1}{\sqrt{\mathrm{det}~\mathbfss{J}}}\exp\left[-\frac{1}{2}\mathbf{\delta t}^T\mathbfss{J}^{-1}\mathbf{\delta t}\right].
\end{equation}

\subsection{Profile Stochasticity}
\label{Section:Stochasticity}

In addition to our jitter model, we also include epoch to epoch variation in the shape of the deterministic profile.  This could be the result of systematic effects, such as errors in polarization calibration, or due to intrinsic shape changes in the profile, where the radiometer noise for a particular observing system is below the level of the intrinsic stochasticity.

We define this stochasticity in two regimes, high and low frequency fluctuations in the profile shape, where by frequency, we refer to the scale of the fluctuations in phase space.

\subsubsection{High frequency stochasticity}

Any stochasticity in the profile that occurs on scales of less than the bin width in phase space will present itself in any one epoch simply as an increase in the uncorrelated noise in the on pulse region of the profile.  In the most general case, we can define a `stochastic envelope', which represents  the increase in the radiometer noise due to this process such that:

\begin{equation}
\mathcal{E}_i(t, \mathbf{\theta_e}, \mathbf{\epsilon}) = \sqrt{\frac{N_i}{T_i}} C_is_p(t, \mathbf{\theta_e}, \epsilon),
\end{equation}
where $\mathbf{\theta_e}$ are the parameters that describe the shape of the envelope, $T_i$ is the integration time for the observation at epoch $i$, and $C_i$ is the normalisation constant defined in Eq. \ref{Eq:normalisation}.  We then add this stochastic profile to our instrument noise $\sigma_i$ such that the new variance in a bin $j$ for observational epoch $i$ is given by:

\begin{equation}
\hat{\sigma}_{ij}^2 = (\alpha_i\sigma_{ij})^2 + \mathcal{E}_i(t, \theta_e, \epsilon)_j^2.
\end{equation}
While this is the most general case, such that the shape of the envelope is free to vary in the analysis, we will assume a simpler model so that the stochastic envelope has the same shape as the mean profile, and is proportional to the amplitude of the profile in any one observation.  We note that giant pulses observed in both canonical pulsars, and MSPs (e.g. \cite{1984bens.work...63W,  1999ApJ...517..460S, 2001ApJ...557L..93R}) show structure on $\sim 10$ns scales, and have been observed to be localised in phase space. In these cases we would expect that our simple model for high frequency profile stochasticity would break down. In principle, however, one could model such fluctuations using a non-Gaussian probability density function (e.g. \cite{2014MNRAS.444.3863L}), or by folding in prior information about the structure of such pulses, though this is not an approach we consider here.

We therefore define a scaling parameter $\beta$ which represents the increase in the variance, and redefine $\mathcal{E}$ such that:

\begin{equation}
\mathcal{E}_i(t, \mathbf{\theta}, \mathbf{\epsilon}) = \beta\sqrt{\frac{N_i}{T_i}}C_is_p(t, \mathbf{\theta}, \epsilon).
\end{equation}

\subsubsection{Low frequency stochasticity}
\label{Section:LowFreqStoc}

To model low-frequency stochastic shape changes in the profile we use the same shapelet basis as for the mean profile itself.  Our goal is then to robustly determine the power spectrum of the shape changes as a function of the scale (or order) of the component in the model.

We therefore define the set of shape variation power spectrum coefficients $\mathbf{\mathcal{S}}$, such that for a given scale $i$ the covariance matrix, which we denote $\mathbfss{S}$,  will be given by:

\begin{equation}
\label{Eq:jitterPrior}
S_{ij} = \left< (\zeta_i - \bar{\zeta}_i)(\zeta_j - \bar{\zeta}_j)\right> = \frac{1}{P_{d,i}}\mathcal{S}_i\delta_{ij},
\end{equation}
where $\bar{\zeta}_i$ is the mean value of the shapelet coefficient $\zeta_i$, which in our analysis we take to be the values that describe the deterministic profile. The normalising factor $1/P_{d,i}$, with $P_{d,i}$ the power in the on pulse region at epoch $i$, is included to give us the power spectrum coefficients $\mathcal{S}_i$ in units of the fraction of the total power in the profile.

In any one observational epoch $i$, we can therefore write the sum of the deterministic, and stochastic components of the profile as:

\begin{equation}
\label{Eq:StocShapelet}
s'_i(t, \mathbf{\bar{\theta}}, \epsilon, \mathbf{\zeta}) = \bar{s}_i(t, \mathbf{\bar{\theta}}, \epsilon) + s_i(t, \mathbf{\theta}, \epsilon).
\end{equation}

We can then write our joint probability density for the individual jitter amplitudes, and the variance parameters as:

\begin{eqnarray}
\label{Eq:Pulsarstoclike}
\mathrm{Pr}(\mathbf{d} | \mathbf{\bar{\theta}}, \mathbf{\zeta}, \bmath{\epsilon}, \mathbf{\mathcal{S}}) &=& \mathrm{Pr}(\mathbf{d} | \mathbf{\bar{\theta}}, \bmath{\epsilon}, \mathbf{\zeta}) \times \mathrm{Pr}(\mathbf{\zeta} | \mathbf{\mathcal{S}})
\end{eqnarray}
where $\mathrm{Pr}(\mathbf{d} |\mathbf{\bar{\theta}}, \mathbf{\epsilon}, \bmath{\zeta})$ has the same functional form as in Eq. \ref{Eq:TimeLike}, but with the the shapelet profile model being replaced by the sum of the mean, and stochastic profile models as in Eq.~(\ref{Eq:StocShapelet}), and $\mathrm{Pr}(\mathbf{\zeta} | \mathbf{\mathcal{S}})$ is given by:

\begin{equation}
\mathrm{Pr}(\mathbf{\zeta} | \mathbf{\mathcal{S}}) \propto \frac{1}{\sqrt{\mathrm{det}~\mathbfss{S}}}\exp\left[-\frac{1}{2}\mathbf{\zeta}^T\mathbfss{S}^{-1}\mathbf{\zeta}\right].
\end{equation}

\section{Marginalising analytically over the linear parameters}
\label{section:Margins}

Ignoring any additional stochastic parameters, our model for the pulse profile so far contains a set of $n_\mathrm{max}-1$ shapelet amplitudes, which are defined across all observing systems that operate across some bandwidth, and some period of time and describe the shape of the mean profile.   At any observational epoch however, effects such as scintillation in the interstellar medium can cause both the effective band-integrated shape, and the amplitude of the profile to vary significantly \citep{1969Natur.221..158R}.  As such, the overall amplitude must be free to vary from one epoch to another, and in addition, the baseline offset must also be treated as a separate parameter for each epoch.  

When including stochastic parameters we then have an additional amplitude parameter per epoch that describes the shift in the arrival time due to pulse jitter, as defined in Section \ref{Section:Jitter}, as well as a set of amplitude parameters that describe the low frequency stochastic variation in the pulse shape for that epoch as described in Section \ref{Section:LowFreqStoc}.

In principle this introduces a large number of additional free parameters, however we can marginalise analytically over all of these amplitude parameters without introducing large matrix operations and thus significantly decrease the size of parameter space we must sample over numerically.  

If the number of amplitude parameters we wish to marginalise over analytically for any epoch $i$ is $N_\mathrm{m}$ we define a $N_\mathrm{m}\times N_i$ matrix, which we denote $\mathbfss{P}_\mathbf{i}$ such that, for example the profile overall amplitude parameter and offset amplitude would be included as:

\begin{eqnarray}
(P_i)_{1,j} &=& 1 \nonumber \\
(P_i)_{2,j} &=& (s_i)_j, \nonumber \\
\end{eqnarray}
with the remaining basis vectors describing pulse jitter and profile variation being included in rows $(P_i)_{3..N_\mathrm{m}}$. We then define the diagonal, $N_\mathrm{m} \times N_\mathrm{m}$ matrix $\mathbf{\Psi_i}$, which combines our priors for the jitter and low frequency stochastic parameters, and is zero for the elements corresponding to the overall amplitude and baseline offset.   Finally we define the length-$N_\mathrm{m}$ vector $\mathbf{b_i}$, which contains the amplitude parameters that multiply our basis vectors in $\mathbfss{P}_\mathbf{i}$, which allows us to write our likelihood as:

\begin{eqnarray}
\label{Eq:TimeLike2}
\mathrm{Pr}(\mathbf{d} |\mathbf{\bar{\theta}}, \mathbf{\epsilon}, \mathbf{\delta t}, \mathbf{\zeta})\mathrm{Pr}(\mathbf{\zeta} | \mathbf{\mathcal{S}})\mathrm{Pr}(\mathbf{\delta t} | \mathbf{\mathcal{J}}) &\propto& \prod_{i=1}^{N_d} \frac{1}{\sqrt{\mathrm{det}\mathbf{N_i}\mathrm{det}\mathbf{\Psi_i}}} \times\\
& & \hspace{-1cm}\exp{\left[-\frac{1}{2}(\mathbf{d}_\mathbf{i} - \mathbfss{P}_\mathbf{i}\mathbf{b_i})^T\mathbfss{N}_\mathbf{i}^{-1}(\mathbf{d}_\mathbf{i} - \mathbfss{P}_\mathbf{i}\mathbf{b_i})\right]} \times \nonumber \\
& & \exp{\left[-\frac{1}{2}\mathbf{b_i}^T\mathbf{\Psi_i}^{-1}\mathbf{b_i}\right]} \nonumber.
\end{eqnarray}

In order to perform the marginalisation over the amplitudes $\mathbf{b_i}$, we define  $\mathbfss{P}_\mathbf{i}^T\mathbfss{N}_\mathbf{i}^{-1}\mathbfss{P}_\mathbf{i} + \mathbf{\Psi_i}^{-1}$ as $\mathbf{\Sigma}_\mathbf{i}$ and $\mathbf{P}_\mathbf{i}^T\mathbf{N}_\mathbf{i}^{-1}\mathbf{d}_\mathbf{i}$ as $\mathbf{\bar{d}}_\mathbf{i}$ and then integrate over the elements in $\mathbf{b}_\mathbf{i}$ analytically to get the likelihood that the data is described by the remaining parameters alone.


Our marginalised probability distribution for the remaining parameters alone is then given by:

\begin{eqnarray}
\label{Eq:MarginAmp}
\mathrm{Pr}(\mathbf{\bar{\theta}}, \mathbf{\epsilon}, \mathbf{\mathcal{S}}, \mathbf{\mathcal{J}} | \mathbf{d}) &\propto& \prod_{i=1}^{N_d} \frac{\mathrm{det} \left(\mathbf{\Sigma}_\mathbf{i}\right)^{-\frac{1}{2}}}{\sqrt{\mathrm{det}\left(\mathbfss{N}_\mathbf{i}\right)}} \nonumber \\
&\times& \exp\left[-\frac{1}{2}\left(\mathbf{d}_\mathbf{i}^T\mathbfss{N}_\mathbf{i}^{-1} \mathbf{d}_\mathbf{i} - \mathbf{\bar{d}}_\mathbf{i}^T\mathbf{\Sigma}_\mathbf{i}^{-1}\mathbf{\bar{d}}_\mathbf{i}\right)\right]. 
\end{eqnarray}

\section{Bayesian analysis methods}
\label{Bayesian}

Bayesian Inference provides a consistent approach to the estimation of a set of parameters $\Theta$ in a model or hypothesis $\mathcal{H}$ given the data, $D$.  Bayes' theorem states that:

\begin{equation}
\mathrm{Pr}(\Theta \mid D, \mathcal{H}) = \frac{\mathrm{Pr}(D\mid \Theta, \mathcal{H})\mathrm{Pr}(\Theta \mid \mathcal{H})}{\mathrm{Pr}(D \mid \mathcal{H})},
\end{equation}
where $\mathrm{Pr}(\Theta \mid D, \mathcal{H}) \equiv \mathrm{Pr}(\Theta)$ is the posterior probability distribution of the parameters,  $\mathrm{Pr}(D\mid \Theta, \mathcal{H}) \equiv L(\Theta)$ is the likelihood, $\mathrm{Pr}(\Theta \mid \mathcal{H}) \equiv \pi(\Theta)$ is the prior probability distribution, and $\mathrm{Pr}(D \mid \mathcal{H}) \equiv Z$ is the Bayesian Evidence.

For model selection, the evidence is key, and is simply the factor required to normalise the posterior over $\Theta$:

\begin{equation}
Z = \int L(\Theta)\pi(\Theta) \mathrm{d}^n\Theta,
\label{eq:Evidence}
\end{equation}
where $n$ is the dimensionality of the parameter space.

As the average of the likelihood over the prior, the evidence is larger for a model if more of its parameter space is likely and smaller for a model where large areas of its parameter space have low likelihood values, even if the likelihood function is very highly peaked.  Thus, the evidence automatically implements Occam's razor: a simpler theory with a compact parameter space will have a larger evidence than a more complicated one, unless the latter is significantly better at explaining the data.

The question of model selection between two models $\mathcal{H}_0$ and $\mathcal{H}_1$ can then be decided by comparing their respective posterior probabilities, given the observed data set $D$, via the posterior odds ratio $R$:

\begin{equation}
R= \frac{P(\mathcal{H}_1\mid D)}{P(\mathcal{H}_0\mid D)} = \frac{P(D \mid \mathcal{H}_1)P(\mathcal{H}_1)}{P(D\mid \mathcal{H}_0)P(\mathcal{H}_0)} = \frac{Z_1}{Z_0}\frac{P(\mathcal{H}_1)}{P(\mathcal{H}_0)},
\label{Eq:Rval}
\end{equation}
where $P(\mathcal{H}_1)/P(\mathcal{H}_0)$ is the {\it a priori} probability ratio for the two models, which can often be set to unity but occasionally requires further consideration.

The posterior odds ratio then allows us to obtain the probability of one model compared with the other, simply as:

\begin{equation}
P = \frac{R}{1+R}.
\end{equation}

Typically in our analysis we deal with the log odds ratio, which is then simply the difference in the log Evidence for two competing models. Table \ref{Table:EvidenceDefs} lists a common interpretation of the Bayes factor given by \cite{bayesRef}.  In all our analysis we will consider a difference in the log Evidence between models of 3 to be the minimum required to justify the addition of extra model components.

\begin{table}
\caption{Interpretation of the evidence}
\begin{tabular}{cccc}
\hline\hline
$R$               &    $\log (R)$       &	P		&    Strength of evidence          \\
\hline
$1 \to 3$	  &	$0 \to 1$	&	$0.5\to0.75$	& 	Not worth more than a bare mention\\
$3 \to 20$	  &	$1 \to 3$	&	$0.75\to0.95$	&	Positive\\
$20 \to 150$      &	$3 \to 5$	&	$0.95\to0.99$	&	Strong\\
$>150	$         &	$> 5$		&	$>0.99$		&	Very strong\\
\hline
\end{tabular}
\label{Table:EvidenceDefs}
\end{table}

\subsection{PolyChord}

\begin{figure*}
\begin{center}$
\begin{array}{cc}
\includegraphics[width=90mm]{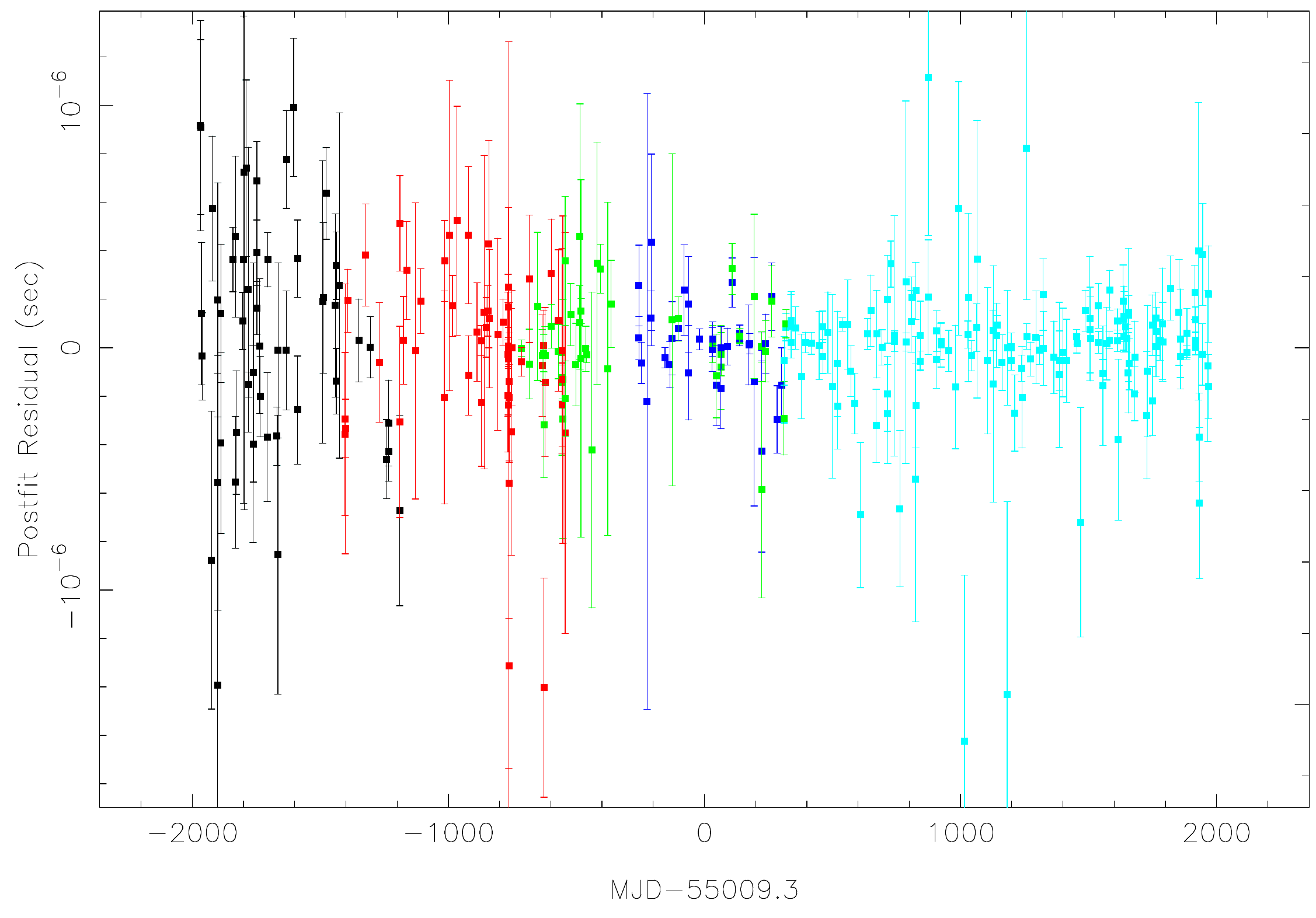} &
\includegraphics[width=90mm]{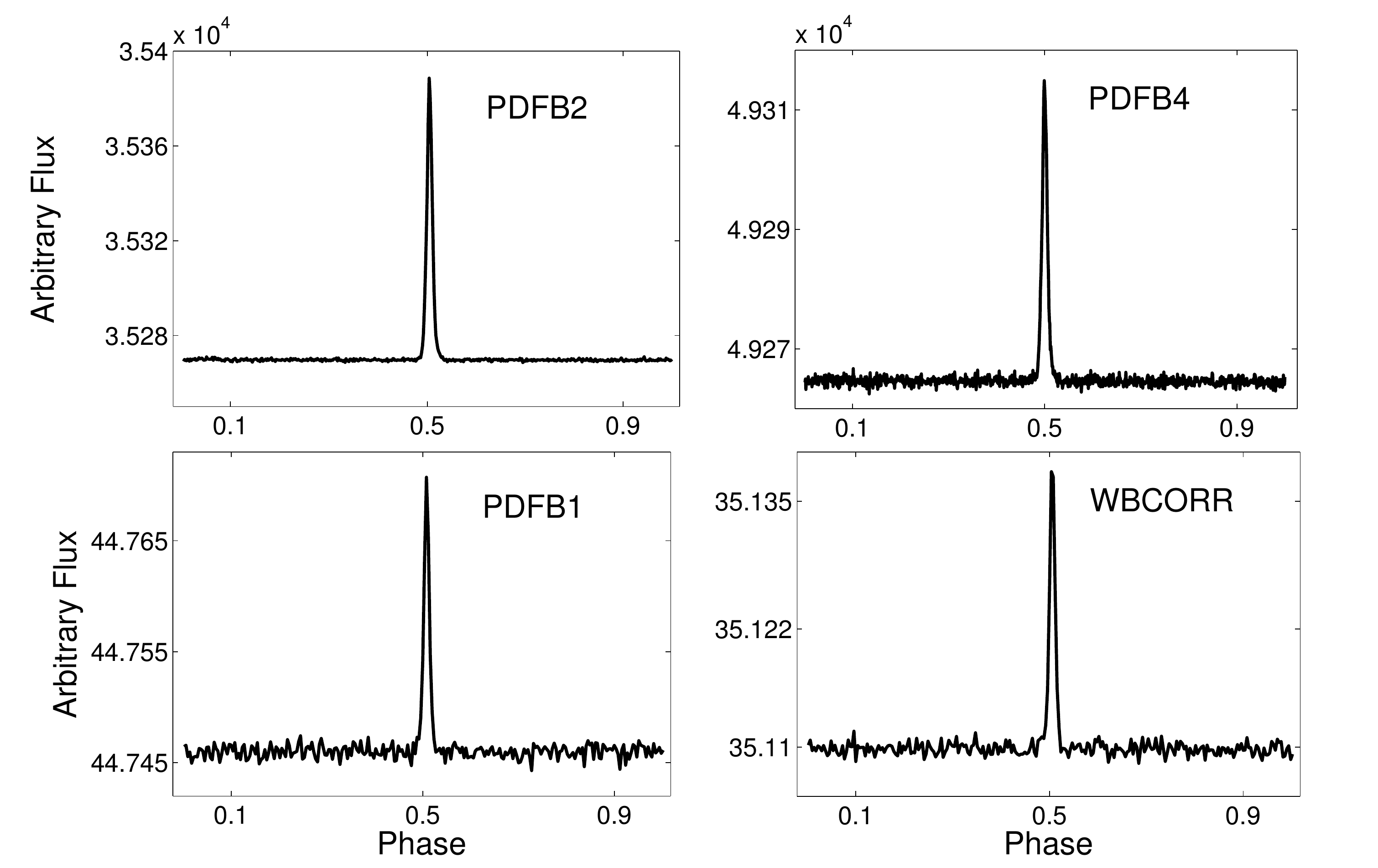}\\
\end{array}$
\end{center}
\caption{(Left) Post-fit residuals for PSR J1909$-$3744 after subtracting the maximum likelihood timing model without including any additional stochastic parameters in the TOA domain analysis.  The weighted RMS is 112ns with a reduced $\chi^2$ of 1.7.  Colours represent different observing systems: (black) WBCORR, (red) PDFB1, (green) PDFB2, (dark blue) PDFB4, (light blue) PDFB4a. (Right) Example profile data for PSR J1909$-$3744 from four different observational epochs, each with a different observing system.  The different overall fluxes, offsets and noise levels in each are apparent.}
\label{Fig:J1909Res}
\end{figure*}

The nested sampling approach \citep{2004AIPC..735..395S}  is a Monte-Carlo method targeted at the efficient calculation of the evidence, but also produces posterior inferences as a by-product. Recently a new nested sampling algorithm, \textsc{Polychord}  \citep{2015arXiv150201856H}, was introduced that makes evidence calculation in hundreds of dimensions a tractable process.  The \textsc{Polychord}   algorithm makes use of `slice sampling' \citep{2000physics...9028N}.   In one dimension, given a likelihood $L_0$, a point $x$ is within the `slice' if $L(x) > L_0$.  Starting from a seed point $x_0$, sampling boundaries are set by expanding a random initial bound of width $w$ until $L(x_0 \pm w) < L_0$.  A new point $x_1$ is then obtained within the slice by sampling uniformly within these bounds. If $x_1$ is not in the slice, $w$ is decreased and $x_1$ is drawn again from these new bounds.

In $d$ dimensions, \textsc{Polychord}   first `whitens' the parameter space, performing a linear skew transformation which turns degenerate contours in the original parameter space into into ones with dimensions O(1) in all directions.  An initial live point is then chosen randomly with coordinates in this whitened space given by $\mathbf{x_0}$.  A random initial direction is then chosen $\mathbf{\hat{n}_0}$, and one dimensional slice sampling is performed in that direction to generate a new point $\mathbf{x_1}$.  This process is repeated $O(d)$ times to generate a new uniformly sampled point $\mathbf{x_N}$ which is decorrelated from the initial point $\mathbf{x_0}$.

In the analysis presented in this work  we will be dealing with models that have up to $\sim 100$ free parameters, as such \textsc{Polychord} is a vital tool for performing robust model comparison in the profile domain.

\section{Dataset}
\label{Section:Dataset}

We perform our analysis, and construct simulations that are based off the 10cm data set for PSR J1909$-$3744 that was previously presented in S15.  Fig. \ref{Fig:J1909Res} shows the post-fit residuals for the 322 TOAs in this data set, observed with a total of 4 systems over the course of 10.8 yr.

These data were recorded at a centre frequency of $3100$~MHz, with $512$-$1024$~MHz of bandwidth.  Over the 10.8 yr of observation, data were recorded with a number of different spectrometers.  The improved computing power of the later spectrometers enabled the implementation of polyphase digital filters, and an increased number of spectral channels and pulse phase bins.  These data were processed using procedures described in \cite{2013PASA...30...17M}, which we will describe in brief below, in order to draw comparison with the methods presented in this work.

Individual observations were first averaged fully in both time and frequency to form a single pulse profile for each observation.  Pulse TOAs were then measured by convolving model templates in the Fourier Domain with the total intensity pulse profiles for each observational epoch.  Model templates were constructed by summing a series of von Mises functions, each of which is described by 3-parameters, a phase, an amplitude, and a width.  Initial guesses for component locations, amplitudes and widths were added by eye and then fitted using a non-linear algorithm. The residuals of the fit were inspected to determine if the fit was good or if additional components were needed, and the processes iterated upon until the residuals appeared to have white-noise characteristics.

Because of its relatively simple morphology \citep{2015MNRAS.449.3223D} only three components were needed to model the pulse profile, however, as in this work, the weak interpulse was not included in the model.  Templates  were formed for each set of systems that had markedly different backend architectures, in order to deal with profile distortions that might arise from the time, and frequency sampling for the different systems.
Pulse-profile features narrower than a pulse bin will be unresolved.  Similarly, dispersive smearing can broaden a pulse if frequency channels are wide (and, like the systems used here, coherent dedispersion is  not employed) relative to the intra-channel dispersion, however, consistent timing solutions were found if one template was used for all the backends, and at 10 cm this pulsar is not affected by dispersive smearing.

Between backends discrete phase offsets (referred to as jumps) occur because of different propagation times through the digital systems.    These offsets can also occur within a backend when it, or the observatory clock-distribution system experiences significant hardware changes.  While it is possible in principle to measure relative offsets between backends independent of the pulsar observation, (by for example injecting a common signal into multiple backends and measuring the delay), these systems have yet to demonstrate the accuracy necessary to correct the highest-precision timing observations.   As a result, we include as free parameters in our analysis offsets between all of the backends, and, for the Parkes Digital Filterbank (PDFB) 4, at MJD 55319  when hardware changes re-defined the location within the signal processing at which a time-stamp was assigned, resulting in the definition of the PDFB4a system after this jump date.

\section{Simulations}
\label{Section:Simulations}

\begin{figure*}
\begin{center}$
\begin{array}{cc}
\includegraphics[width=90mm]{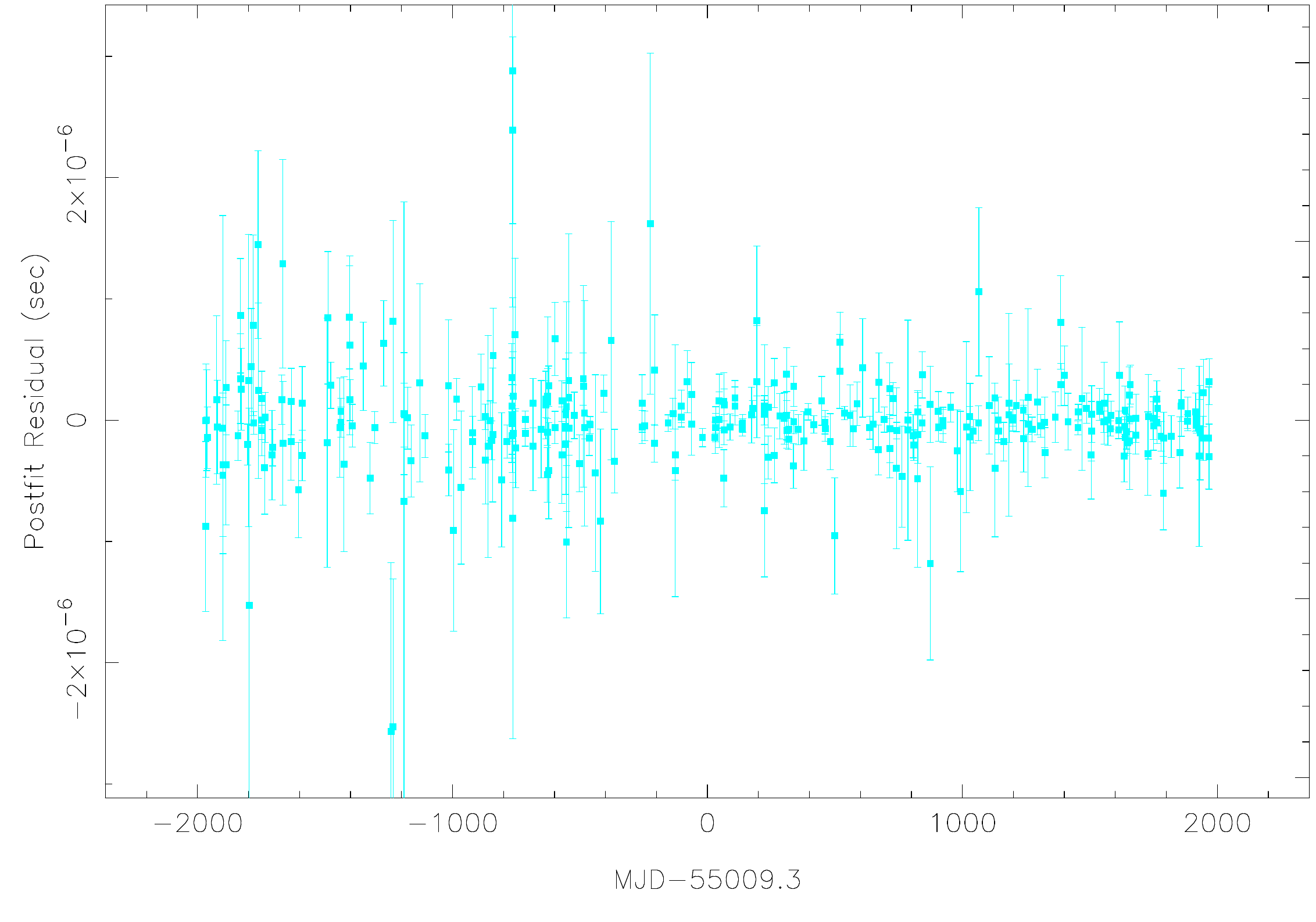} &
\includegraphics[width=90mm]{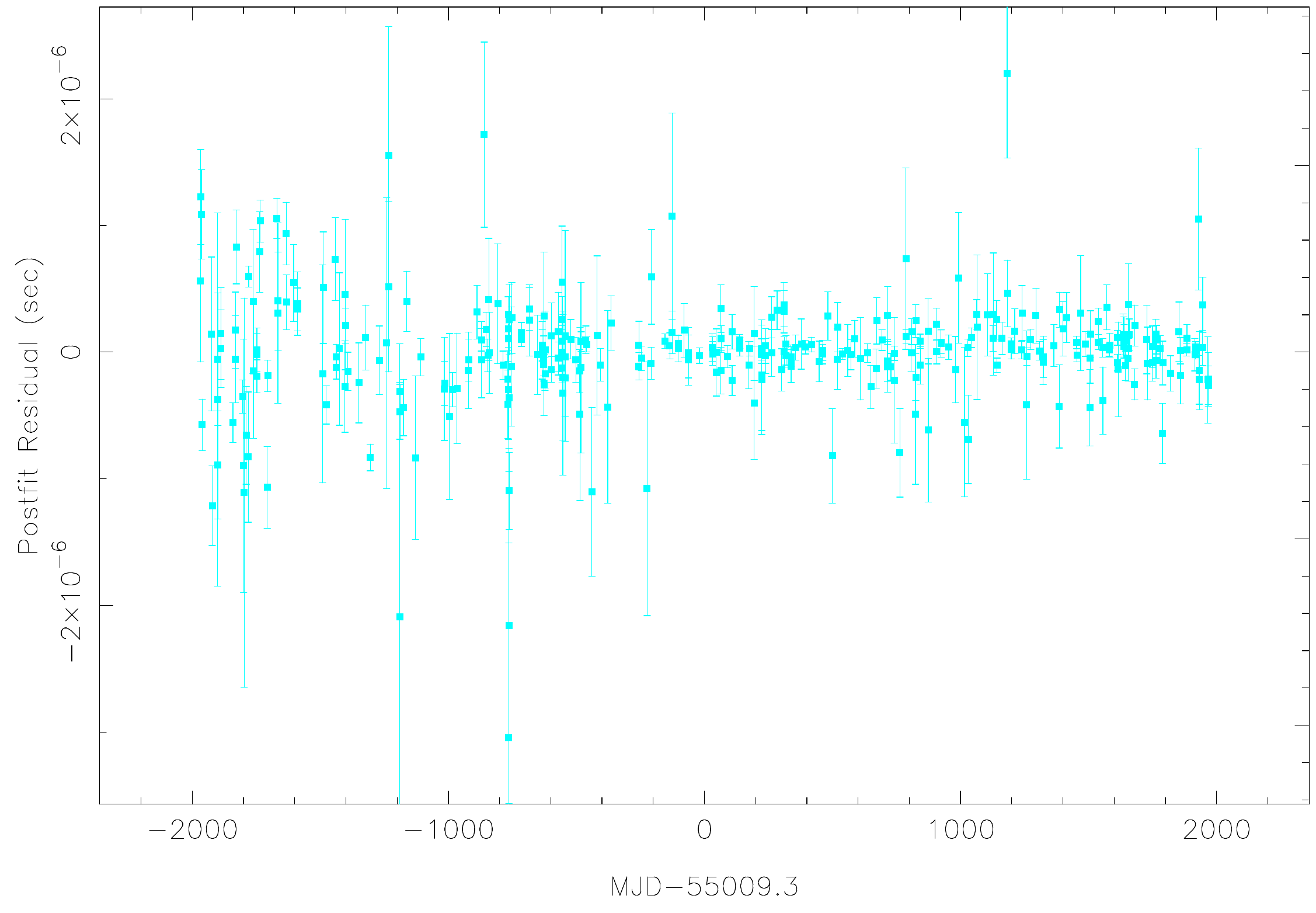} \\
\includegraphics[width=90mm]{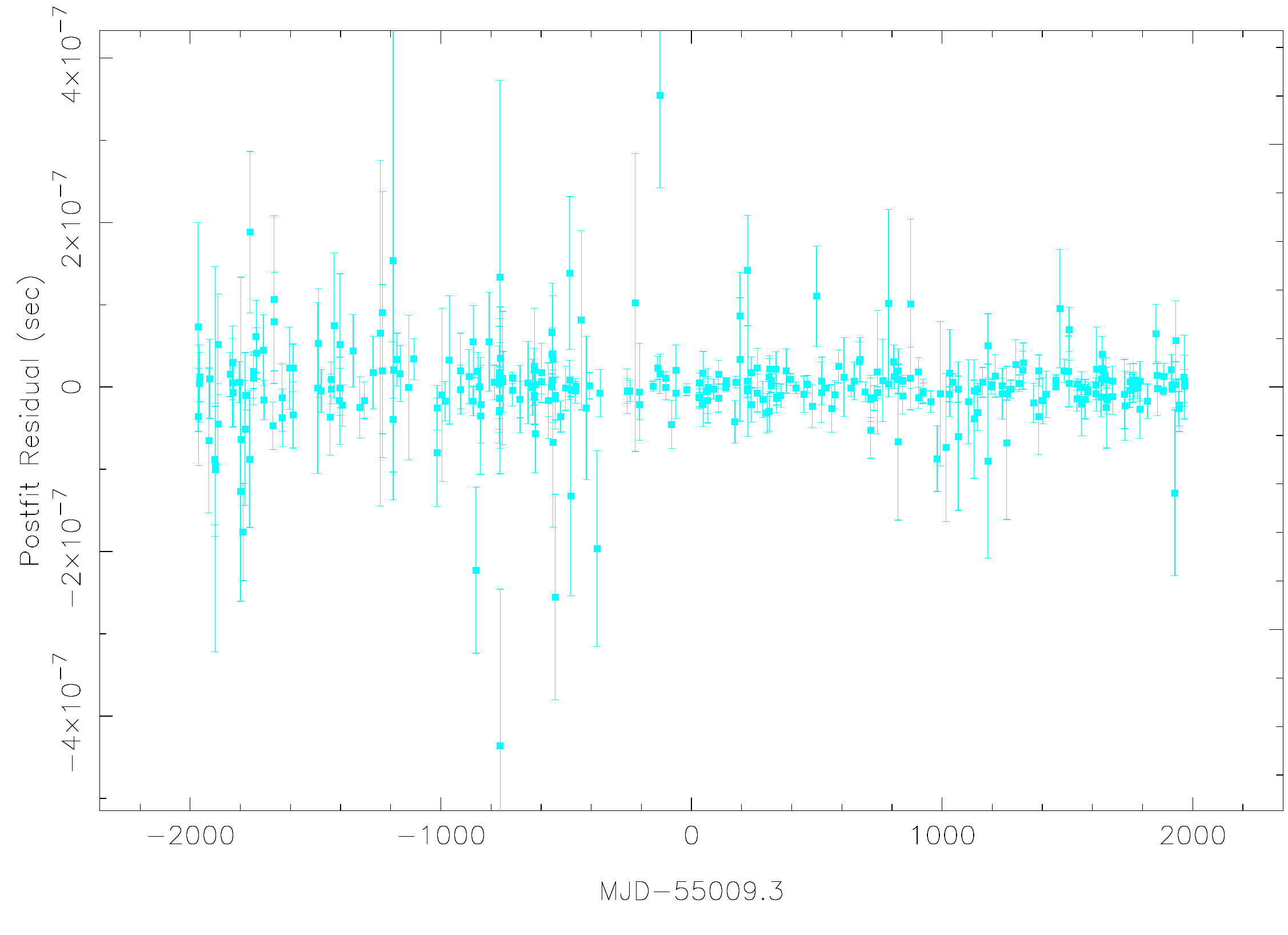} &
\includegraphics[width=90mm]{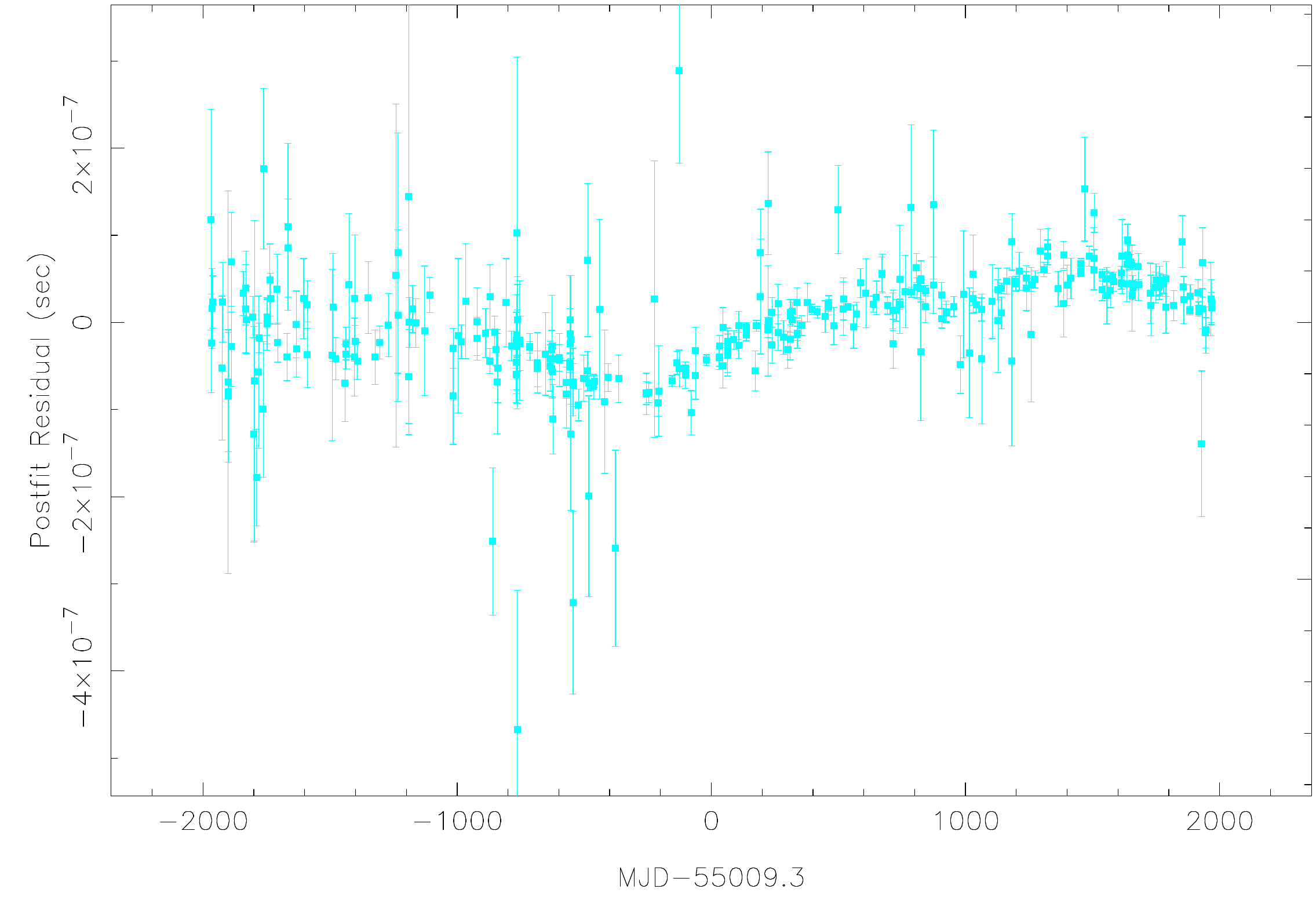} \\
\end{array}$
\end{center}
\caption{Post-fit residuals for PSR J1909$-$3744 simulation 1 (top left), 2 (top right), 3 (bottom left), and 4 (bottom right)  after subtracting the maximum likelihood timing model from a TOA domain analysis in each case. Simulations 1 and 2 include radiometer noise consistent with the levels determined from the real data set, and in simulation 2 we include an additional systematic jitter term to the WBCORR profiles with an amplitude of $5\times10^{-7}$~s, chosen to be consistent with the value observed in the real data set when performing a TOA domain analysis.   Simulations 3 and 4 include a factor 10 times less radiometer noise than the real data set, and in simulation 4 we include an additional Gaussian component to the profile, that has a width $0.1\%$ the pulse period, $1\%$ the amplitude of the profile for any given observational epoch, and which moves deterministically through the main pulse profile over the course of 4000 days.}
\label{Fig:J1909SimRes}
\end{figure*}

We construct four simulations using a fiducial timing model for PSR J1909$-$3744 consistent with published values (e.g. \cite{2011MNRAS.414.3117V}).  In all cases we use a simple Gaussian model for the profile, and use the data set described in Section \ref{Section:Dataset} to determine the observed flux at each epoch, as well as the baseline offset.  We then include different levels of the radiometer noise, systematic pulse jitter, and additional non-stationary profile components to the simulations to test the efficacy of our profile domain jitter and stochasticity models described in Sections \ref{Section:Jitter} and \ref{Section:Stochasticity}. For each simulation we process the profile data in the same way as described in Section \ref{Section:Dataset} to form TOAs against which to compare our profile domain analysis.  

In simulations 1 and 2 we include radiometer noise consistent with the levels determined from the real data set.   In simulation 2 we then include an additional systematic jitter term to the Wide Band Correlator (WBCORR, \citealt{2013PASA...30...17M}) profiles with an amplitude of $5\times10^{-7}$~s, chosen to be consistent with the value observed in the real data set when performing a TOA domain analysis.   In simulations 3 and 4 we include a factor 10 times less radiometer noise than the real data set, and in simulation 4 we include an additional Gaussian component to the profile, that has a width $0.1\%$ the pulse period, and $1\%$ the amplitude of the profile for any given observational epoch.  The component moves deterministically through the main pulse profile over the course of 4000 days.  We use this for our simulation rather than including stochasticity in the same basis as our model in order to check the efficacy of our method when it is attempting to correct for epoch to epoch profile variations that are not well described by zero mean Gaussian fluctuations in the amplitude of the profile components.

In Fig. \ref{Fig:J1909SimRes} we show the residuals for all simulations after subtracting the fiducial timing model.  The additional jitter in the WBCORR system is apparent in the early data for simulation 2, as is the timing noise induced by the deterministic passage of an additional Gaussian component through the pulse profile for simulation 4.

\begin{figure*}
\begin{center}$
\begin{array}{cc}
\includegraphics[width=90mm]{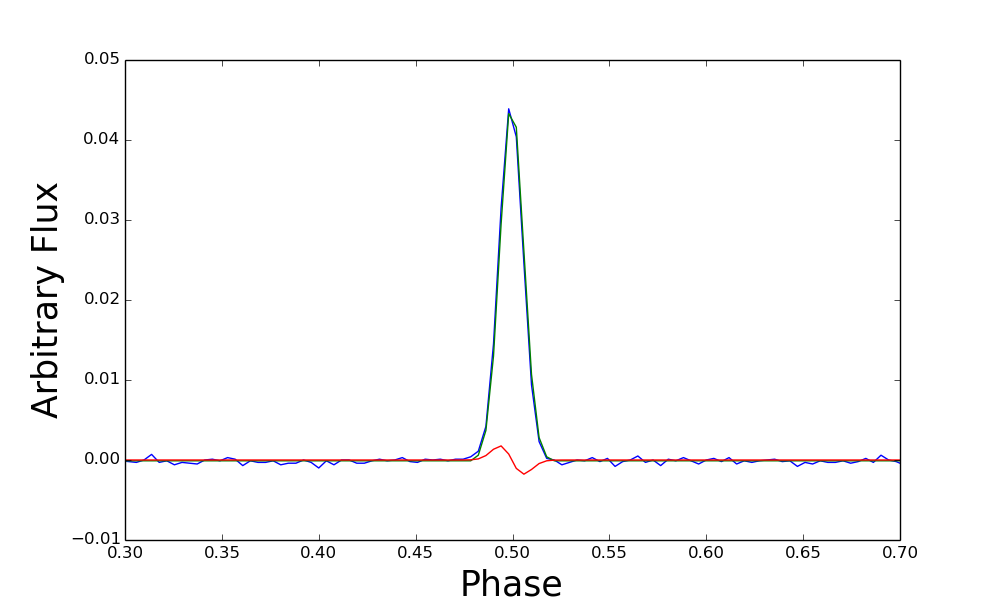} &
\includegraphics[width=90mm]{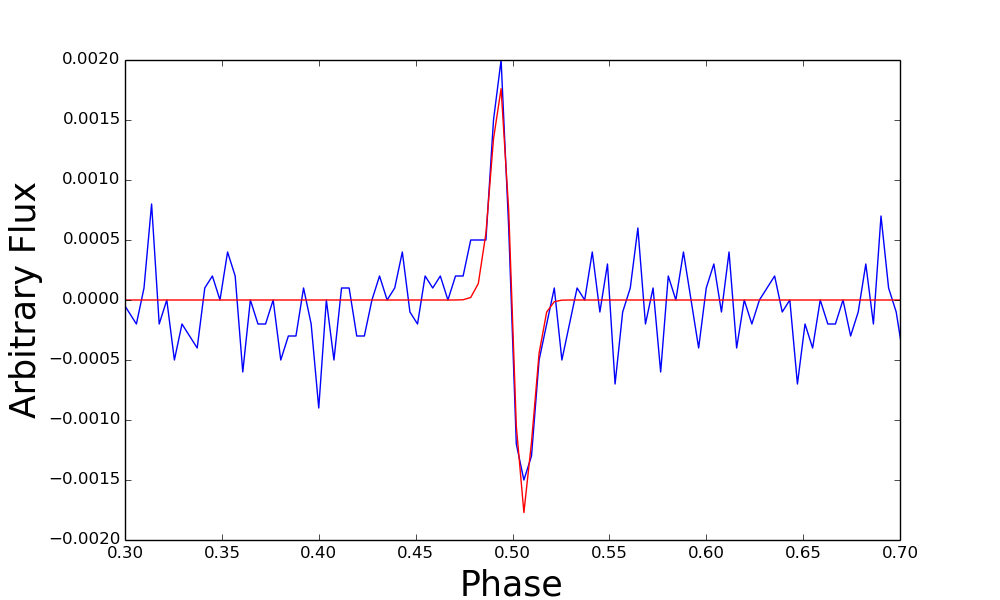} \\
\includegraphics[width=90mm]{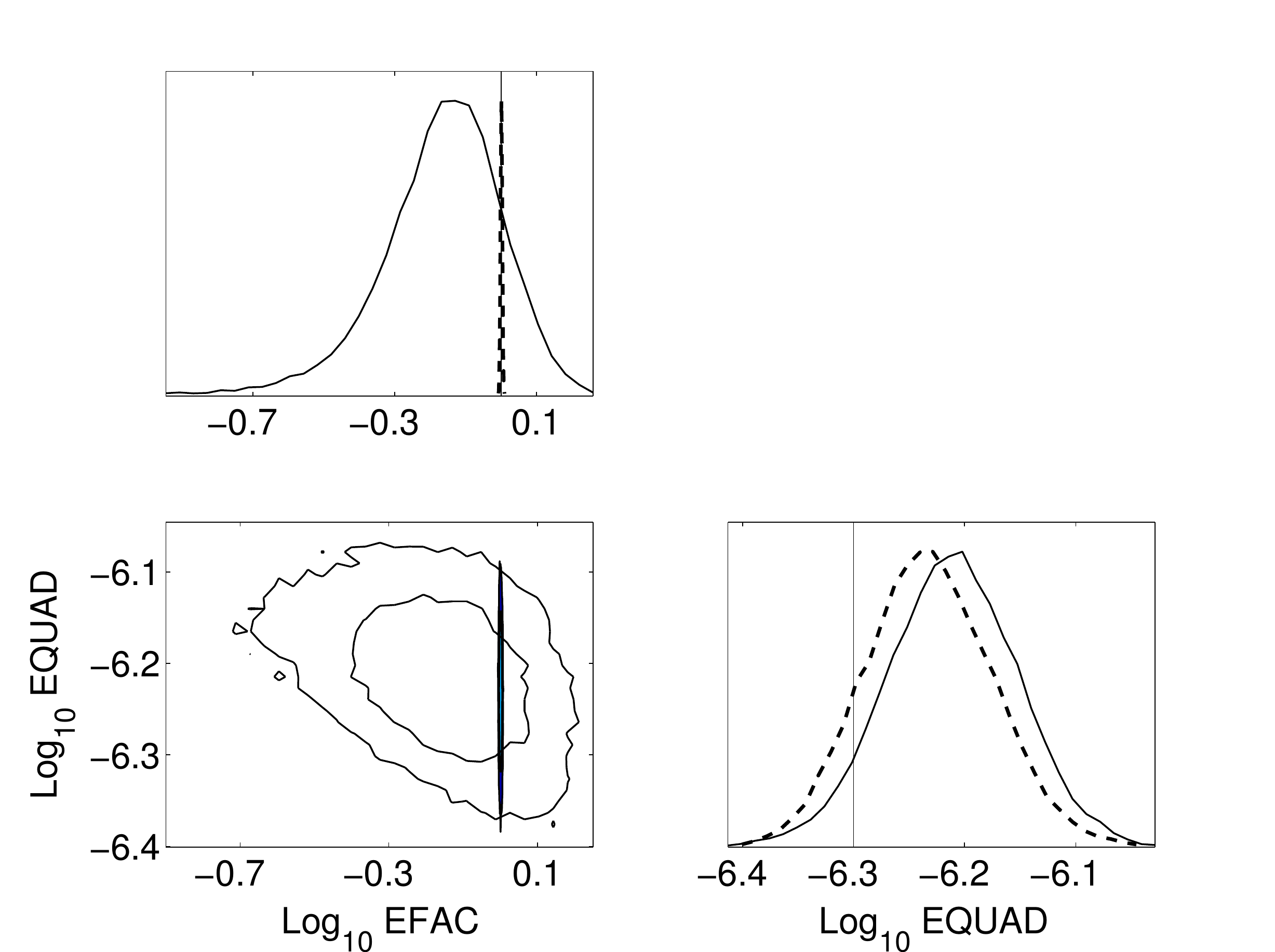} &
\includegraphics[width=90mm]{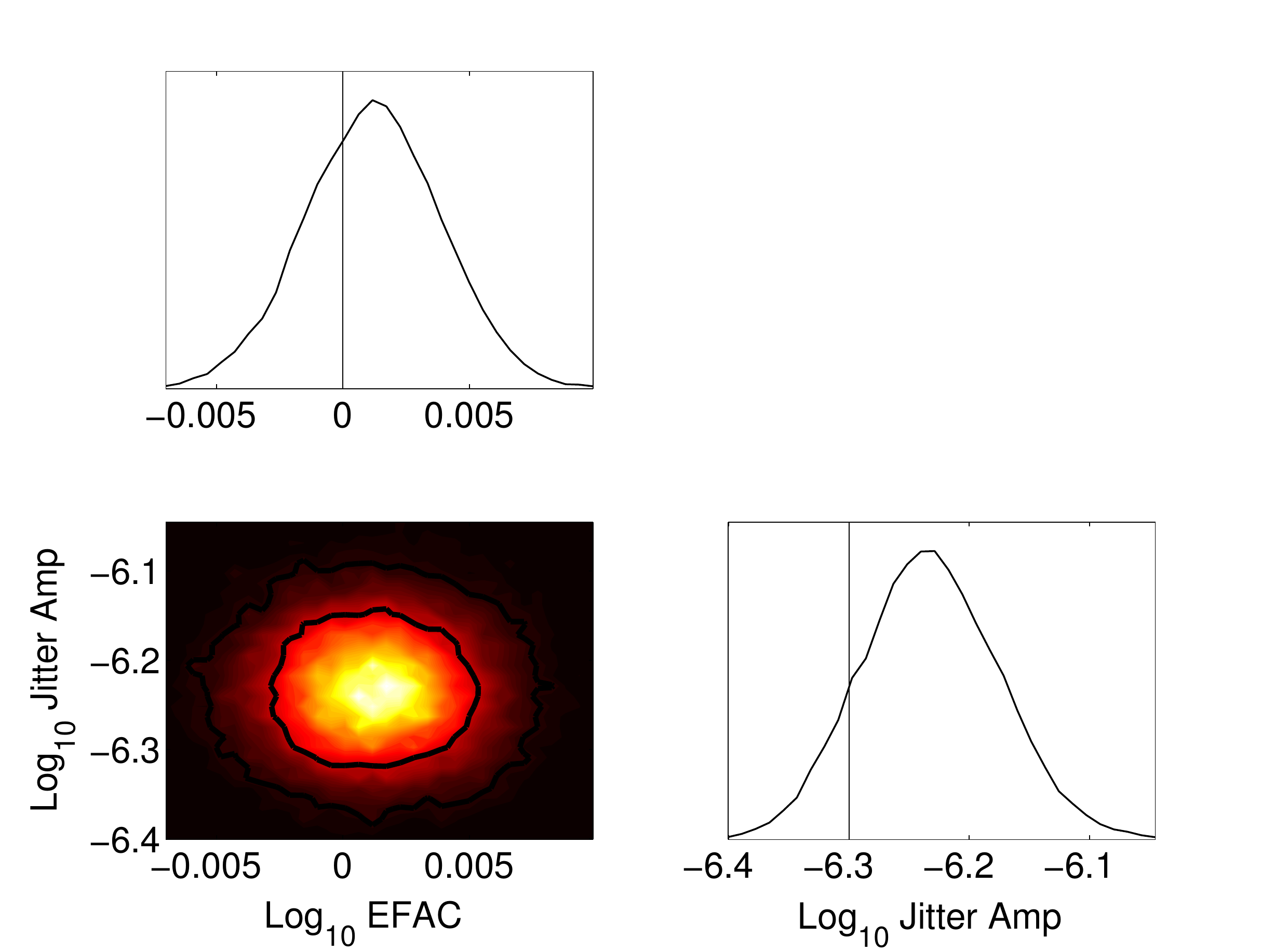} \\
\end{array}$
\end{center}
\caption{(Top left) Simulated profile (blue line) for a single observational epoch from simulation 2, for the WBCORR system, to which we applied the additional jitter signal.  Over plotted are the maximum likelihood model profile, evaluated at the arrival time predicted by the maximum likelihood timing model from our analysis (green line), and the maximum likelihood jitter profile (red line).  (Top right)  As in the previous panel, however we have subtracted the model profile from the data, leaving a residual that is clearly consistent with our jitter model.  (Bottom left) One and two-dimensional posterior parameter estimates for the EFAC and EQUAD parameters for the WBCORR system obtained in the TOA domain analysis (solid lines in the one-dimensional plots, solid contours in the two-dimensional plot), with the EFAC and jitter parameter estimates from our profile domain analysis (dashed lines in the one-dimensional plots, filled dashed contours in the two-dimensional plot).  The uncertainty in the EFAC parameter in the profile domain is an order of magnitude smaller than in the TOA domain, as we can use the statistical information present in the off pulse region to help constrain our estimates of the radiometer noise. The constraints on the jitter amplitude in the profile domain are also improved compared to the TOA domain EQUAD counterpart.  This is because in the TOA domain, the EFAC and EQUAD parameters are correlated (indicated by the ellipsoidal contours).  In the profile domain this is not the case, as seen in the bottom right panel, the jitter amplitude, and EFAC parameters are completely decorrelated in our profile domain analysis.}
\label{Fig:J1909Sim}
\end{figure*}

\subsection{Simulations 1 and 2}

In order to compare the profile and TOA domain analysis for simulations 1 and 2, we define the following models:

\begin{itemize}
\item TOA  Model 1:  Timing model only.
\item TOA Model 2:  Timing model and additional EFAC and EQUAD parameters for each observing system.
\item Profile Model 1:  Deterministic profile and timing model only.
\item Profile Model 2:  Deterministic profile, timing model and additional EFAC and jitter parameters for each observing system.
\end{itemize}


In both profile domain models we include only the $n=0$ shapelet coefficient in our analysis.  We find that the timing model parameter estimates and uncertainties for equivalent models in both simulations are completely consistent, as we would expect, as the assumptions about the stochastic properties in the profile domain were propagated correctly into the TOA analysis. 

\begin{figure*}
\begin{center}$
\begin{array}{cc}
\includegraphics[width=90mm]{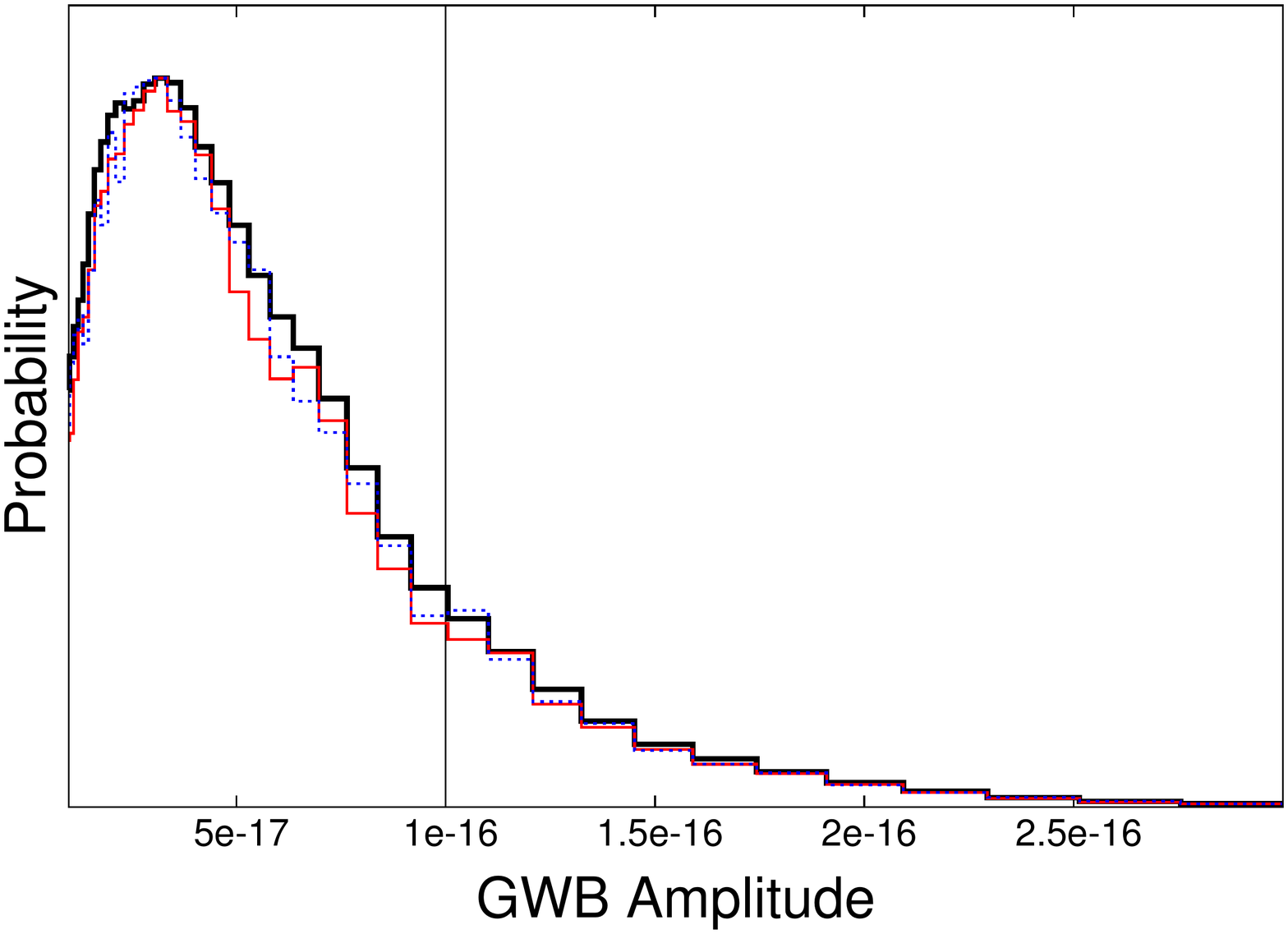} &
\includegraphics[width=90mm]{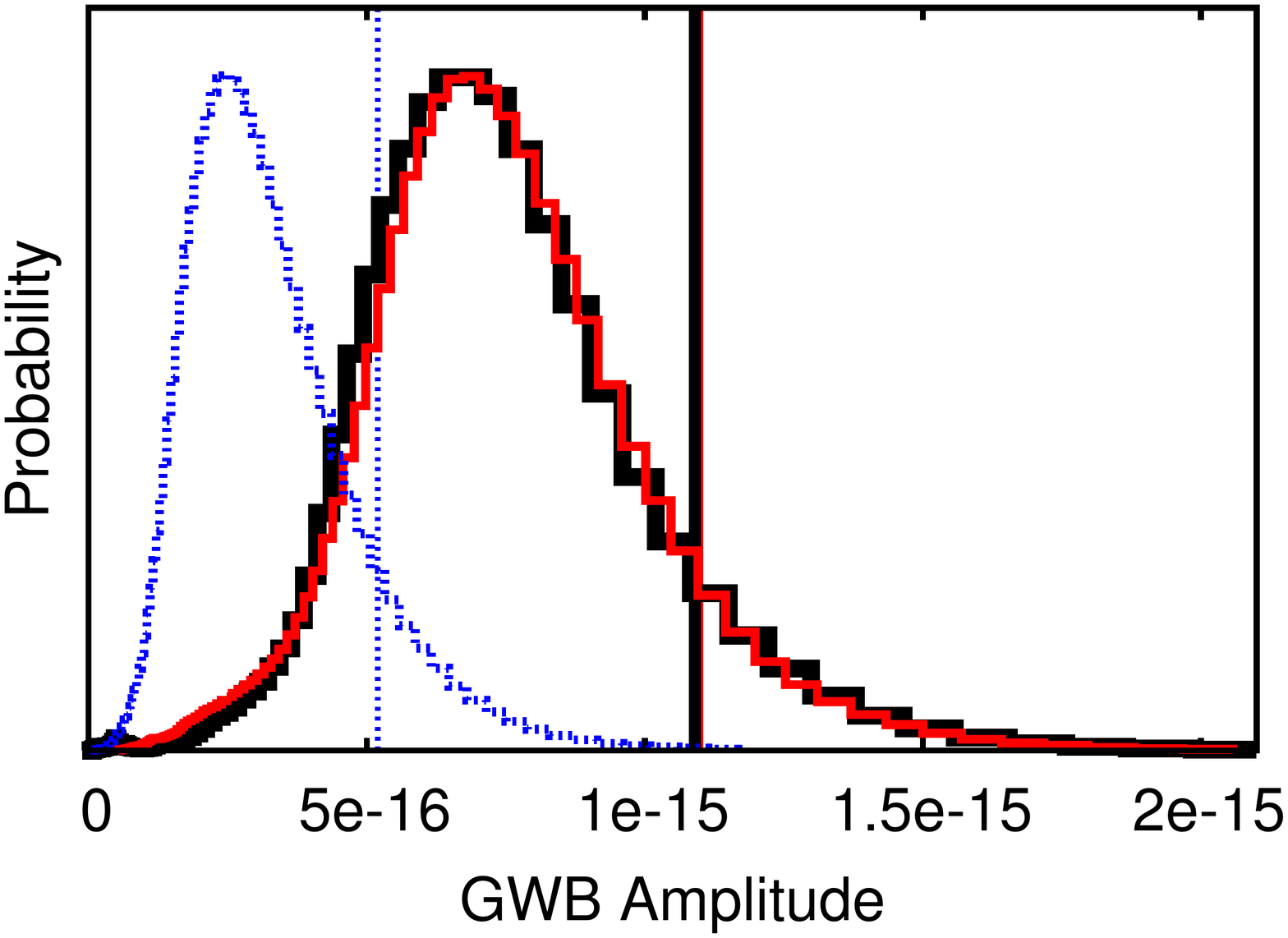} \\
\end{array}$
\end{center}
\caption{One-dimensional marginalised posterior parameter estimates for the amplitude of a steep red noise process with a spectral index of $\gamma = 13/3$ for simulation 3 (left plot) and simulation 4 (right plot).  Both the TOA domain and profile domain analyses for simulation 3 result in consistent 95\% upper limits of $1\times 10^{-16}$.  In simulation 4 we find that the TOA domain analysis, and a profile domain analysis that includes a red noise signal but no profile stochasticity results in consistent 95\% upper limits of $1\times 10^{-15}$.  Despite the fact that the additional Gaussian component was only 1\% of the profile amplitude, and 0.1\% of the pulse width, it has a significant impact on the upper limit, increasing it by an order of magnitude.  Our profile domain analysis that includes profile stochasticity (blue dashed line) significantly improves upon the TOA domain analysis,  with a 95\% upper limit of $5\times 10^{-16}$. While this is still much greater than for the data set that included no profile variation it still demonstrates clearly that by modelling these effects in the profile domain they can be better isolated from the GWB signal of interest. }
\label{Fig:J1909Sim3Limits}
\end{figure*}

In simulation 2 we find that the profile domain jitter model results in superior constraints on the high frequency variation in pulse arrival times compared to the TOA domain EQUAD model.  In Fig. \ref{Fig:J1909Sim} (Top) we show  a simulated profile for a single observational epoch from simulation 2, for the WBCORR system, to which we applied the additional systematic jitter signal.  In the left panel we show the simulated profile (blue line), the model profile, evaluated at the arrival time predicted by the maximum likelihood timing model from our analysis (green line), and the maximum likelihood jitter profile (red line).  In the right panel we have subtracted the model profile from the data, and the residual is clearly consistent with our jitter profile.  

In Fig. \ref{Fig:J1909Sim} (bottom left) we compare the one and two-dimensional posterior parameter estimates for the EFAC and EQUAD parameters for the WBCORR system obtained in the TOA domain analysis (solid lines in the one-dimensional plots, solid contours in the two-dimensional plot), with the EFAC and jitter parameter estimates from our profile domain analysis (dashed lines in the one-dimensional plots, filled dashed contours in the two-dimensional plot).  The most striking feature of this comparison is the difference in constraints placed on the EFAC parameters in both domains.  In the profile domain we still have the off pulse region to constrain our estimates of the radiometer noise, information that is no longer available when we move to the TOA domain, leading to an order of magnitude decrease in precision when the EFAC parameter is estimated as part of the analysis.  As a by product the constraints on the jitter amplitude in the profile domain are improved over the TOA domain EQUAD counterpart.  This is because in the TOA domain, the EFAC and EQUAD parameters are correlated (indicated by the ellipsoidal contours).  In the profile domain this is not the case, as seen in the bottom right panel, the jitter amplitude, and EFAC parameters are completely decorrelated, as a shift in the arrival time of the profile does not {\it look} like an increase in the radiometer noise, however in the TOA domain, a multiplicative prefactor (EFAC), and an additional term added in quadrature (EQUAD), can be completely covariant depending on how similar the error bars are across different observations.

\subsection{Simulations 3 and 4}

For simulations 3 and 4 we consider the following models:

\begin{itemize}
\item TOA Model:  Timing model, EQUAD parameters for each observing system, and additional red timing noise.
\item Profile Model 1:  Deterministic profile, timing model and additional red timing noise.
\item Profile Model 2:  Deterministic profile, timing model, high and low frequency stochastic profile parameters,  and additional red timing noise.
\end{itemize}

Additionally, for both simulations and for all 3 models we obtain upper limits on the amplitude of a steep red noise process with a spectral index of $\gamma = 13/3$, consistent with that expected for a gravitational wave background formed from a population of inspiralling super-massive black hole binaries, \citep{2001astro.ph..8028P}.

As for simulations 1 and 2, we find that for both simulations 3 and 4 that the timing model parameters are consistent between all 3 models, indicating that, as for the jitter parameters included in simulations 1 and 2, including the stochastic profile parameters in our analysis does not adversely affect the precision that we can achieve when no such stochasticity is present in the data.  In addition, in Fig. \ref{Fig:J1909Sim3Limits} (left plot) we find the limits obtained on a steep red noise process within the data set are also consistent between the three models, with 95\% upper limits of $A < 1\times10^{-16}$ in all cases.  

For simulation 4, however, we find that the profile domain analysis results in a factor 2 improvement in upper limit when including profile stochasticity, compared to either the standard TOA domain analysis or the equivalent analysis using Model 1 in the profile domain.  We show this explicitly in  Fig. \ref{Fig:J1909Sim3Limits} (right plot), in which we plot the upper limits for the same 3 models applied to simulation 4.  While TOA Model 1 (black line), and Profile Model 1 (red line) are consistent, with 95\% upper limits of $A < 1\times10^{-15}$, we find a factor 2 improvement in the upper limit when incorporating profile stochasticity in our analysis (blue line), obtaining $A < 5\times10^{-16}$.  The only difference between simulations 3 and 4 was the inclusion of an additional non-stationary Gaussian component to the profile model, that moved through the profile over a period of 4000 days.  

Despite having an amplitude of only  $1\%$ that of the profile in any observational epoch, the effect on our TOA domain analysis was to increase the upper limit by an order of magnitude. While our stochastic profile domain model does not fully recover the upper limit obtained from simulation 3, we purposefully simulated non-stationary behaviour in the pulse profile that was not well described by a random Gaussian process.  Despite this, we still obtain a significantly improved limit compared to the TOA domain analysis.  

The key result both from this comparison, and from the results of the jitter analysis in simulations 1 and 2, is that by performing the analysis using the profile data, rather than the SATs as is typical in pulsar timing analysis, we can decorrelate profile variation from the GW signal and significantly improve both our sensitivity to GWs, and our understanding of the stochastic processes present in the data.  As models for profile variation improve, so too will the upper limits on GWs obtainable from a profile domain analysis.  Given the covariance between the signal induced in the timing residuals from GWs, and from long-term temporal profile variation (see Fig. \ref{Fig:MoneyPlot}), however, only by performing a simultaneous fit can robust estimates of both signal components be obtained, without one absorbing the other.

\begin{figure}
\begin{center}$
\begin{array}{c}
\includegraphics[width=90mm]{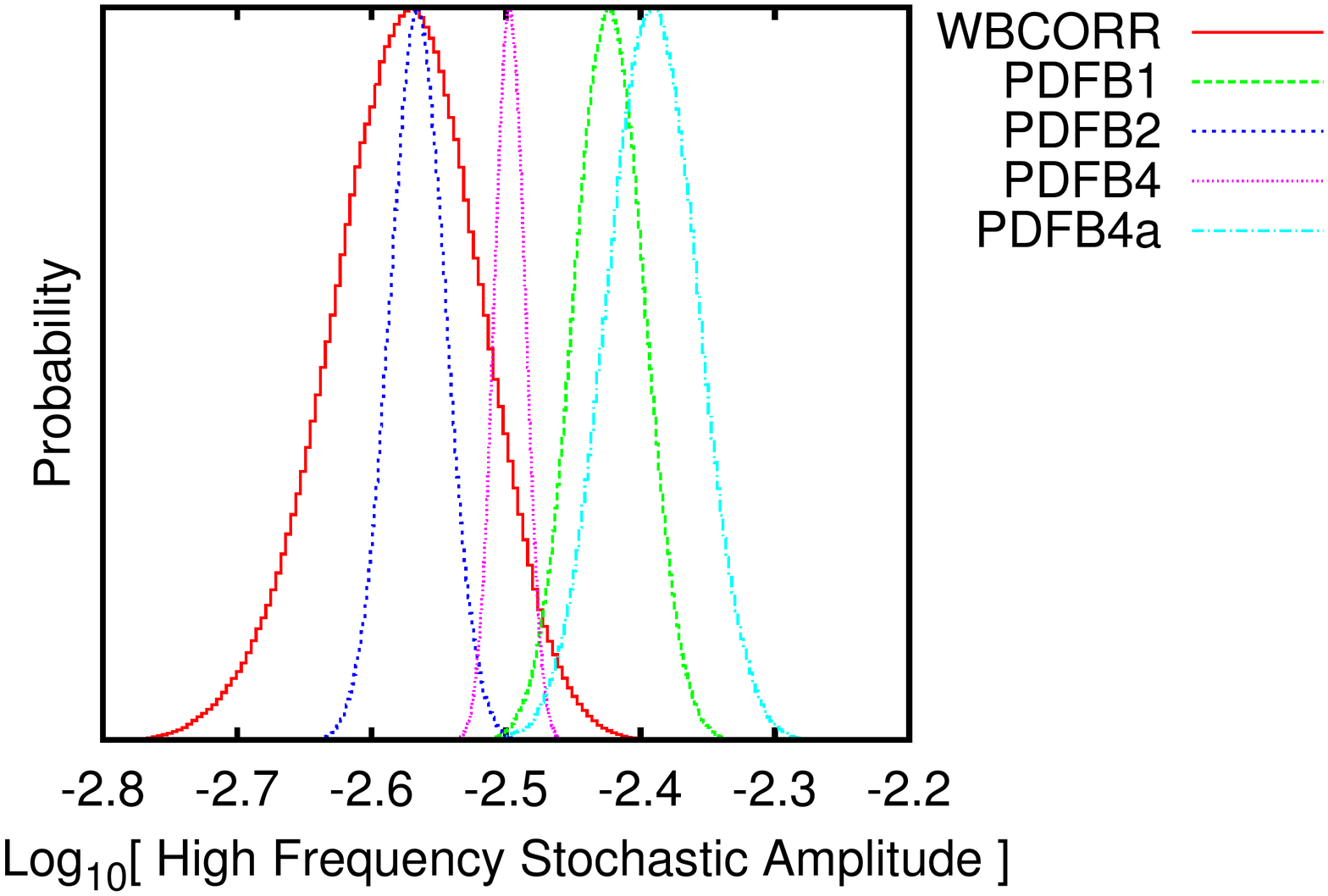} \\
\hspace{-2cm}
\includegraphics[width=90mm]{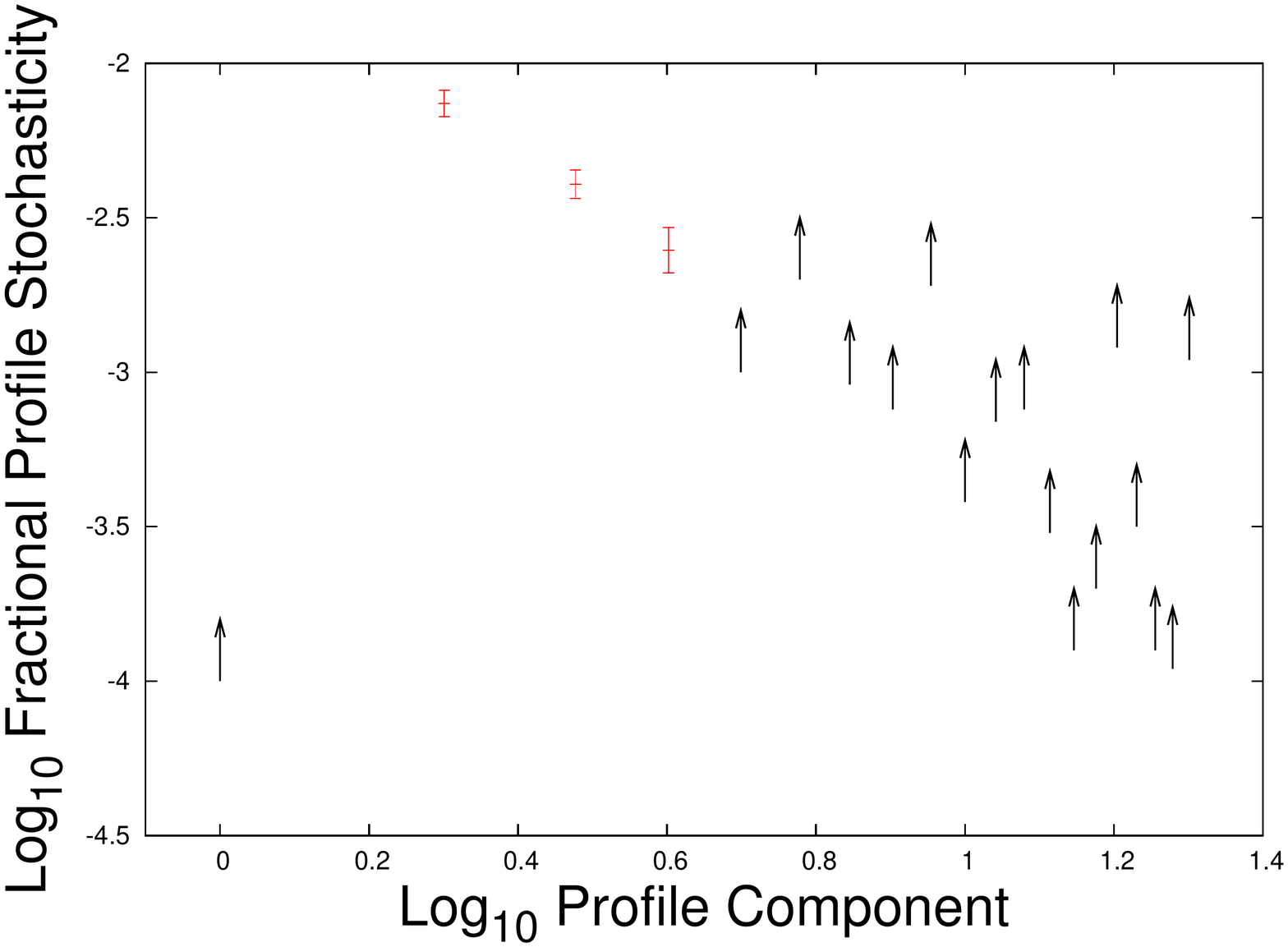} \\
\end{array}$
\end{center}
\caption{(Top) One-dimensional posterior parameter estimates for the amplitude of the high frequency stochastic element to the profile for Simulation 4.  Separate amplitudes were fit to each observing system.  The units of the abscissa are fraction of the total profile amplitude added to the uncorrelated variance for each bin. The amplitudes are consistent with $< 1\%$ of the profile amplitude for all systems however they show a factor $\sim 2$ spread relative to one another.  Given each system samples a different period of time (See Fig. \ref{Fig:J1909Res}) this could be the result of the non-stationary nature of the additional Gaussian component added to the profile data.  (Bottom) Mean values and 1$\sigma$ confidence intervals (red points with error bars), and 95\% upper limits (black arrows) for low frequency stochastic variations in the profile.  Upper limits are plotted where the posterior for that scale is consistent with zero at greater than 5\% probability.}
\label{Fig:J1909Sim4Stochastic}
\end{figure}

We find the amplitude of both the high and low frequency stochastic models are consistent with the properties of the additional profile component added to simulation 4.  In Fig. \ref{Fig:J1909Sim4Stochastic} we show the one dimensional marginalised posterior parameter estimates for both the high frequency (top panel) and low frequency (bottom panel) stochastic profile parameters from our analysis of simulation 4.  We included separate scaling factors, $\beta$, describing the high frequency profile variation for each observing system.  Values are quoted as the fraction of the total profile amplitude added to the uncorrelated variance for each bin. For all systems we find the fractional increase in the uncorrelated noise is $< 1\%$, however the value for different systems varies by a factor $\sim 2$ relative to one another.  Given each system samples a different period of time (See Fig. \ref{Fig:J1909Res}) this could be the result of the non-stationary nature of the additional Gaussian component added to the profile data.

For the low frequency profile variations we show either the mean parameter estimates and 1$\sigma$ confidence intervals (points with error bars), or 95\% upper limits (arrows).  Upper limits are plotted where the posterior for that scale is consistent with zero at greater than 5\% probability.  We find that only stochastic variations in the $n=(2,3,4)$ profile coefficients are significant. Given the width of the profile, $\Lambda = 0.007$ in phase, which corresponds to $\sim 7$ phase bins in the more modern observing systems, beyond $n=4$ the scale of the fluctuations in the stochastic model will be $\le 2$ bins, and so we would expect these to become increasingly covariant with the high frequency stochastic parameters, and thus less significant in our analysis.

\section{Application to the PSR J1909$-$3744 data set}
\label{Section:RealData}

We now apply our profile domain framework to the PSR J1909$-$3744 data set described in Section \ref{Section:Dataset}.  As with the simulations we will compare both TOA and profile domain models, defining:

\begin{itemize}
\item TOA Model 1:  Timing model only.
\item TOA Model 2:  Timing model, and additional EFAC and EQUAD parameters for each observing system.
\item Profile Model 1:  Deterministic profile, and timing model only.
\item Profile Model 2:  Deterministic profile, timing model, and additional EFAC and jitter parameters for each observing system.
\item Profile Model 3:  Deterministic profile, timing model, and additional EFAC, jitter and high frequency stochastic profile parameters for each observing system.
\item Profile Model 4:  Deterministic profile, timing model, EFAC parameters for each observing system and high and low frequency stochastic profile parameters.
\end{itemize}

For each model we find the optimal number of shapelet coefficients to include by increasing the number of included coefficients in steps of 5 until the change in the $\log$ evidence is less than 3.  Table \ref{Table:1909DataEvValues} lists the optimal number of coefficients included in each model, along with the $\log$ evidence values for that model.

\begin{table}
\caption{Evidence values for different models fit to the PSR J1909$-$3744 data set.}
\begin{tabular}{lcc}
\hline\hline
Model & N$_\mathrm{c}$ & $\log \mathcal{Z}$ \\
\hline
TOA Model 1      &  -    &   0       \\
TOA Model 2  	 &  -    &   35.7     \\
\hline 
Profile Model 1  &  40   &   0       \\
Profile Model 2  &  30   &   154.2   \\
Profile Model 3  &  30   &   1302.7  \\
Profile Model 4  &  30   &   1280.5  \\
\hline
\hline
\end{tabular}
\label{Table:1909DataEvValues}
\end{table}

We find that Profile model 1, which includes no additional stochastic parameters requires the largest number of components (40), while subsequent models that include either profile jitter or profile stochasticity require 25$\%$ fewer components to describe the data.  Initially this may seem like a large number compared to the 3 von Mises functions used in the standard analysis, however in this case a separate profile model is fit to the WBCORR, PDFB1, PDFB2, and PDFB4 systems.  Given a set of $n$ von Mises functions require 3$n$-1 parameters ($n$ amplitudes, $n$ widths and $n-1$ relative positions), this means that for the total data set 32 parameters were used to describe the profile shape, which is consistent with our results.

\begin{figure}
\begin{center}$
\begin{array}{c}
\includegraphics[width=80mm]{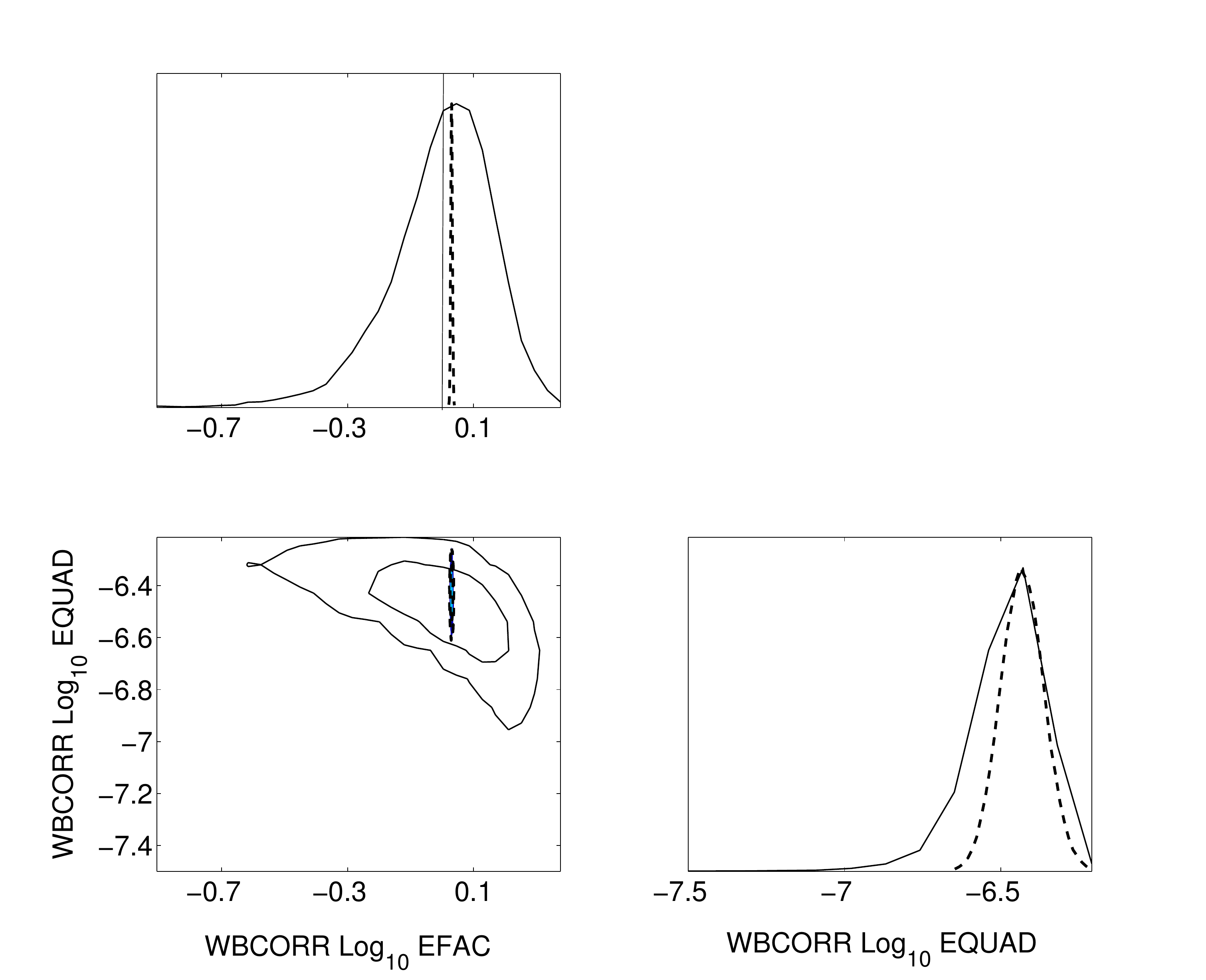} \\
\includegraphics[width=80mm]{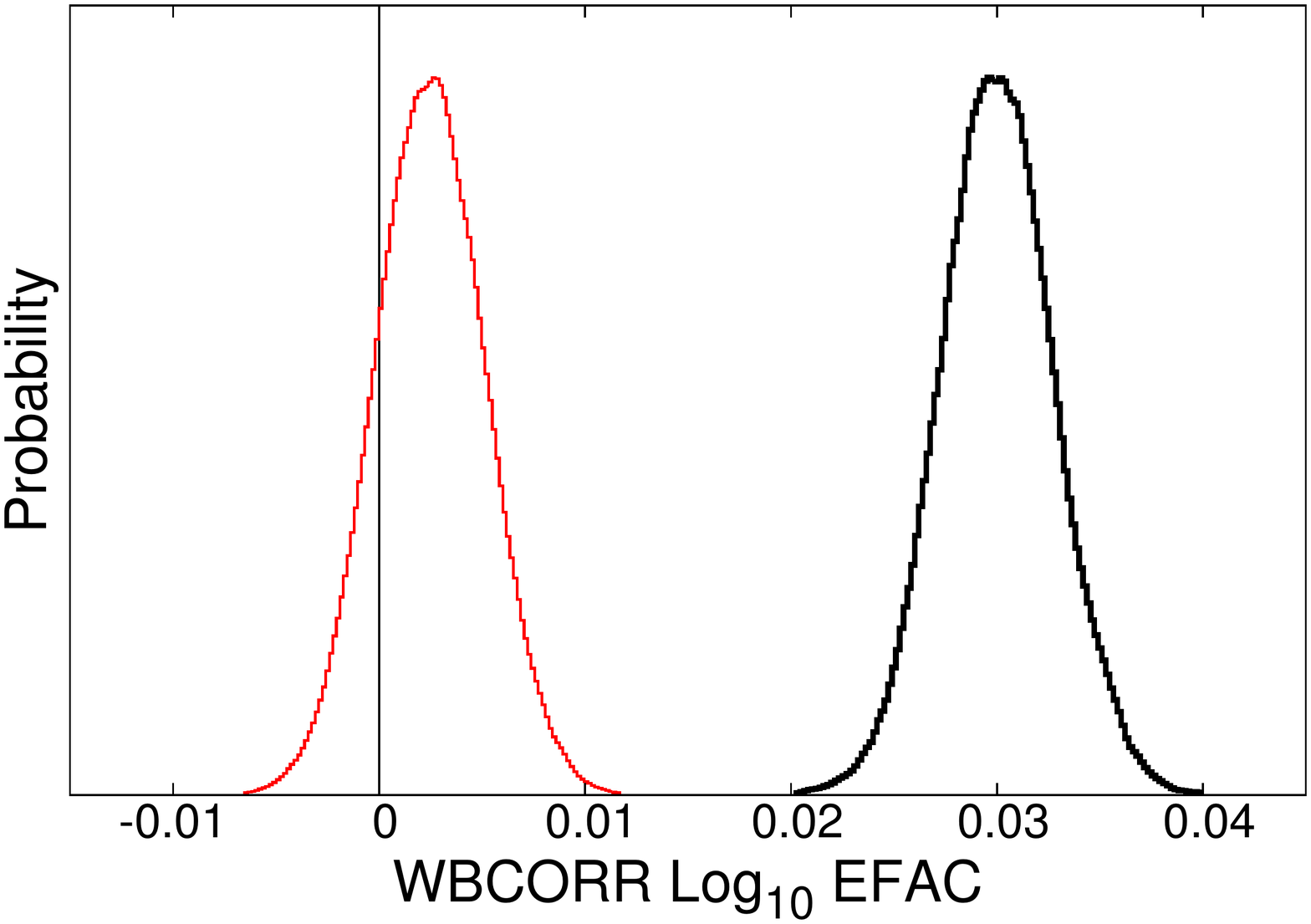} \\
\end{array}$
\end{center}
\caption{(Top) One and two-dimensional posterior parameter estimates for the EFAC and EQUAD parameters for the WBCORR system obtained in the TOA domain analysis (solid lines in the one-dimensional plots, solid contours in the two-dimensional plot), with the EFAC and jitter parameter estimates from our profile domain analysis (dashed lines in the one-dimensional plots, filled dashed contours in the two-dimensional plot).  As in simulation 2, we find that the inferred amplitude of the profile domain jitter model was consistent with the TOA domain EQUAD model.  We similarly see that as before the EFAC parameters are better constrained in the profile domain by approximately an order of magnitude, with no observed correlation with the jitter parameters, leading to better overall constraints being placed on their amplitudes.  (Bottom)  Comparison of the one-dimensional marginalised posterior parameter estimates for the WBCORR EFAC parameter from profile models 2 (black line) and 3 (red line).  Assuming that the mean profile suffers only a shift from epoch to epoch as in profile model 3 (and in the standard timing analysis paradigm) results in an EFAC that is systematically greater than 1, implying that the model used is not sufficient to describe the on pulse region.  This is confirmed by the Bayesian evidence, which is significantly greater ($\Delta \log \mathcal{Z} > 1000$) for a model that includes high frequency, systematic profile stochasticity, which results in an EFAC that is consistent with 1.}
\label{Fig:J1909EFACEQUAD}
\end{figure}

We find timing model parameter estimates and uncertainties consistent between TOA model 1 and profile model 1, and between TOA model 2 and profile models 2-4.  Additionally, as in simulation 2, we find that the inferred amplitude of the profile domain jitter model is consistent with the TOA domain EQUAD model.  Fig. \ref{Fig:J1909EFACEQUAD} (top plot) shows a comparison of the one and two--dimensional posteriors for the TOA EQUAD model (solid lines in the 1d plots, contours in the 2d plot) with the profile domain jitter model (dashed lines in the 1d plots, filled contours in the 2d plot) for the WBCORR system.  As in the simulations we find the EFAC parameters are better constrained in the profile domain by approximately an order of magnitude, and show no observed correlation with the jitter parameters leading to better overall constraints being placed on their amplitudes.  The EFAC parameter that scales the uncorrelated instrumental noise for the WBCORR system, however,  is not consistent with a value of 1 in profile model 3, implying that the residuals are not consistent with our initial estimate of the instrumental noise for this backend.  This can be seen more clearly in the bottom panel of  Fig. \ref{Fig:J1909EFACEQUAD}, which shows a comparison of the one-dimensional marginalised posterior parameter estimates for the WBCORR EFAC parameter from profile models 2 (black line) and 3 (red line).  Assuming that the mean profile suffers only a shift from epoch to epoch as in profile model 3 (and in the standard timing analysis paradigm) results in an EFAC that is systematically greater than 1, implying that the model used is not sufficient to describe the on pulse region.  This is confirmed by the Bayesian evidence, which is significantly greater ($\Delta \log \mathcal{Z} > 1000$) for a model that includes high frequency, systematic profile stochasticity, which results in an EFAC that is consistent with 1.  In addition, when including a high frequency stochastic profile model simultaneously with the jitter parameters as in model 3, we find the significant EQUAD term detected in the TOA domain, and the jitter term seen in profile model 2 is consistent with zero, implying that all of the additional high frequency noise observed in the TOA domain is due to this high frequency stochastic term.

\begin{figure}
\begin{center}$
\begin{array}{c}
\includegraphics[width=80mm]{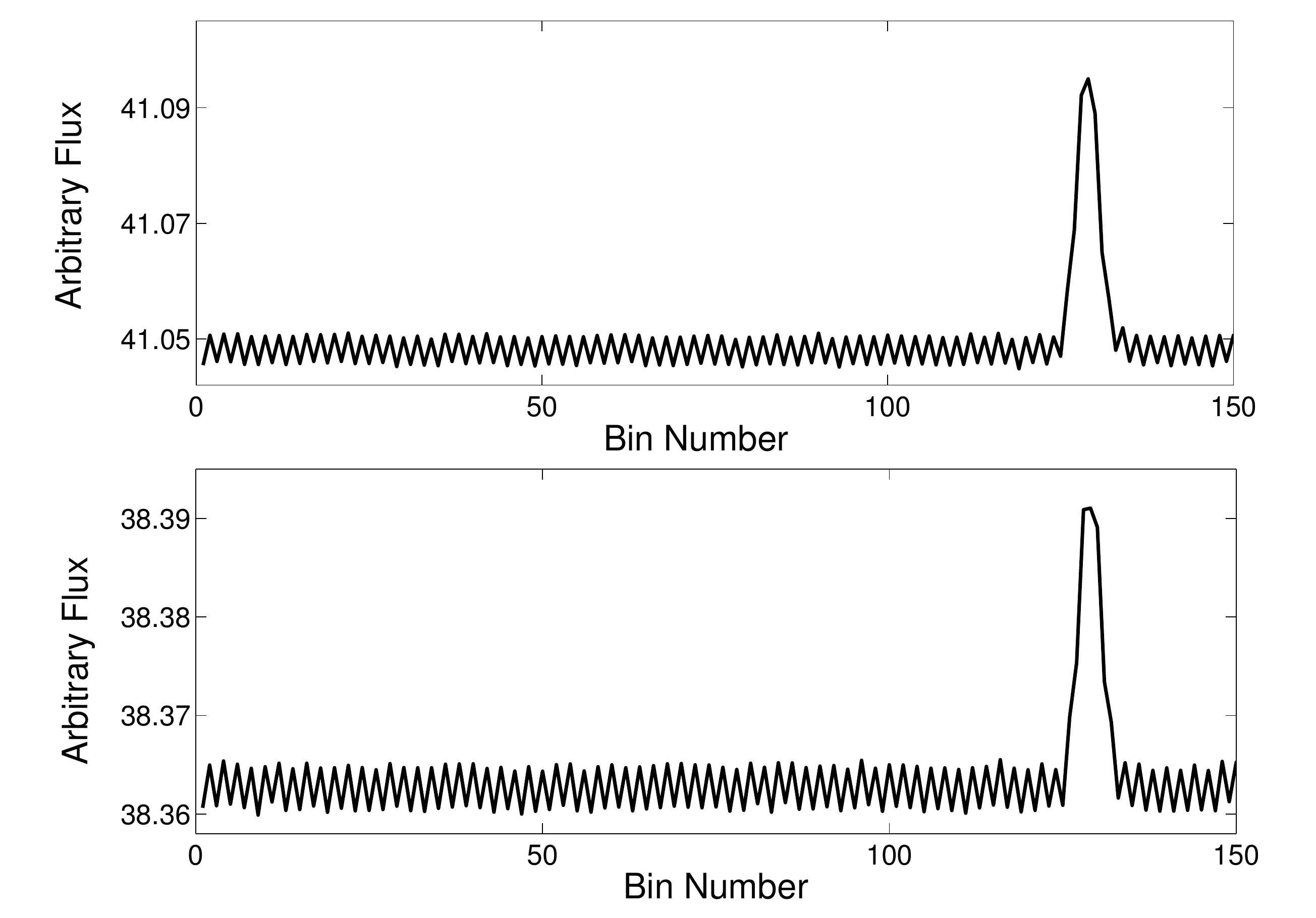} \\
\end{array}$
\end{center}
\caption{Example of two WBCORR profiles that suffer from non-Gaussian `sawtooth' noise.  Approximately 10\% of WBCORR points show this effect by eye, however the amplitude varies from profile to profile. In the TOA domain this manifests itself as an EQUAD term -- high frequency variation in the arrival times of the pulses.  In the profile domain we find the Bayesian evidence is significantly in support ($\Delta \log \mathcal{Z} > 1000$) of a model where the pulse profile in the WBCORR data has a high frequency stochastic component.  When fitting simultaneously for a jitter model, and for a high frequency stochastic element, we found the jitter term to be consistent with zero, suggesting the EQUAD detected in the TOA domain originates in this process.  While clearly the optimal solution would be to simply fix such issues at the level of forming the folded data, where that is not possible by working in the profile domain one can better model such systematic effects and separate them from `true' shifts in the arrival time due to pulse jitter.}
\label{Fig:J1909BadProfiles}
\end{figure}

The origin of this high frequency noise can clearly be seen in Fig. \ref{Fig:J1909BadProfiles}, which shows two example of profiles from the WBCORR system that suffer from non-Gaussian `sawtooth' noise.  Approximately 10\% of WBCORR points show this effect by eye, however the amplitude varies from profile to profile. While clearly the optimal solution would be to simply fix such issues at the level of forming the folded data, where that is not possible by working in the profile domain one can better model such systematic effects and separate them from `true' shifts in the arrival time due to pulse jitter.

\begin{figure}
\begin{center}$
\begin{array}{c}
\includegraphics[width=80mm]{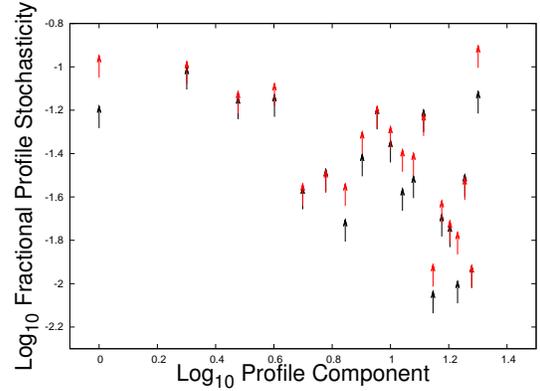} \\
\end{array}$
\end{center}
\caption{95\% upper limits on the fractional profile stochasticity in the PSR J1909$-$3744 data set described in Section \ref{Section:Dataset} obtained using profile model 4.  Red arrows represent upper limits from the full analysis, black arrows represent upper limits when fixing both the timing model, and mean profile model to their maximum likelihood values obtained from profile model 3.  Upper limits range from 1-10\% of the total power in the profile.  When fixing the timing model and mean profile to their maximum likelihood values we find erroneously stringent upper limits by a factor of up to $\sim 1.8$.    }
\label{Fig:1909LFUL}
\end{figure}

Finally, we find no evidence for low frequency stochasticity in the PSR J1909$-$3744 data set using profile model 4.  We show the 95\% upper limits from our joint analysis for the 20 lowest order shapelet components in Fig. \ref{Fig:1909LFUL}, with units given in terms of the fraction of total power in the profile.  Upper limits range from 1-10\% of the total power in the profile.

\subsection{Upper limits on an isotropic stochastic GWB}

\begin{figure}
\begin{center}$
\begin{array}{c}
\includegraphics[width=80mm]{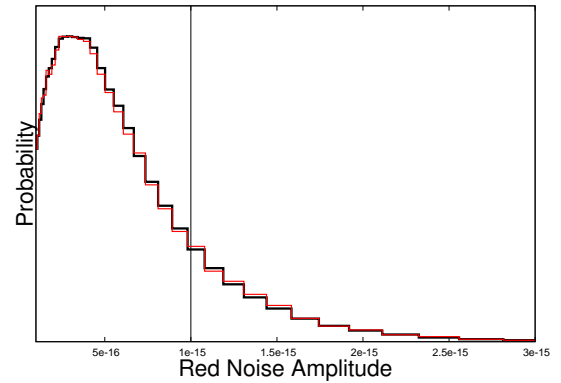} \\
\end{array}$
\end{center}
\caption{One-dimensional marginalised posterior parameter estimates for the amplitude of a red noise process in the PSR J1909$-$3744 data set described in Section \ref{Section:Dataset} at a spectral index of $13/3$ for (black line) a TOA domain analysis using TOA model 2, and (red line) a profile domain analysis using profile model 3.   We find both are consistent with a $95\%$ upper limit of $1\times10^{-15}$, consistent with values reported in S15.  In this data set we found no evidence of low frequency profile stochasticity, and only the older WBCORR system was significantly affected by high frequency systematics, so it is not surprising that upper limits using the optimal models for both the TOA and profile domain analyses are consistent with one another.}
\label{Fig:J1909RealRedUL}
\end{figure}

As a final comparison, we take TOA model 2, and profile model 3 and include a red noise model for each.  In both cases we obtain upper limits on the amplitude of a steep red noise process with a spectral index of $\gamma = 13/3$.  Fig. \ref{Fig:J1909RealRedUL} shows the one-dimensional marginalised posterior parameter estimates for the amplitude of the red noise process for the  TOA domain (black line), and profile domain (red line) analysis.  Both are consistent with one another, with $95\%$ upper limit of $1\times10^{-15}$.  Given that we found no evidence for low frequency profile stochasticity in the data, and only the older WBCORR system, which contributes relatively little weight to the full data set, was significantly affected by high frequency systematics, it is not surprising that upper limits using the optimal models for both the TOA and profile domain analyses are consistent with one another.

\section{Conclusions}
\label{Section:Conclusions}

In this paper we have extended the Bayesian framework `Generative Pulsar Timing Analysis' to incorporate both pulse jitter (high frequency variation in the arrival time of the pulse) and epoch to epoch stochasticity in the shape of the pulse profile.  This framework allows for a full timing analysis to be performed using the folded profile data, rather than the site arrival times as is typical in most timing studies.  

We first used simulations to show that while in a standard TOA domain analysis of pulsar timing data the effect of profile stochasticity and the signal induced by a gravitational wave background are highly covariant, these two signals can be decorrelated when performing the analysis in the profile domain, resulting in a significantly improved upper limit on a stochastic gravitational wave background. We showed that this was also true for high frequency variation in the arrival time of the pulse.  In the TOA domain this is typically modelled using an `EQUAD' parameter, that adds in quadrature to the TOA uncertainties.  This is highly covariant with a scaling of the error bars (using 'EFACs'), which is typically used to model uncertainty in the value of the radiometer noise.  In the profile domain a model for pulse jitter is completely decorrelated from the model for the thermal radiometer noise, resulting in improved constraints on the magnitude of the effect in the pulsar timing data.

Our simulation included temporal variation in the pulse profile in the form of an additional Gaussian component with an amplitude 1\% that of the main profile, and a width 0.1\% of the pulse period.  Despite the relatively small amplitude of the additional signal, this resulted in an order of magnitude increase in the upper limit obtained in the TOA domain analysis.  The techniques described in this paper allow for the simultaneous evaluation of both contributions to the data, resulting in robust upper limits, and eventual detection of a GWB.

We also applied our methodology to a PSR J1909$-$3744 10~cm data set and find significant evidence for systematic high-frequency profile variation resulting from non-Gaussian instrumental noise in the oldest observing system in the analysis.  We find that the evidence supports this description of the data over a jitter model where the variation is modelled simply as a shift of the arrival time of the mean profile.   We find no evidence for either detectable pulse jitter, or low-frequency profile variation, and set limits on the latter of between 1 and 10\% for the different profile model components.  Using our profile domain framework we obtain upper limits on a red noise process with a spectral index of $\gamma = 13/3$ of $1\times10^{-15}$, consistent with previously published limits.

With current upper limits for a stochastic GWB now beginning to rule out existing models for black hole binary populations, profile variability at the $< 1\%$ level can become an important limiting factor in the sensitivity achievable in pulsar timing experiments.  By moving away from the standard timing paradigm, where a timing model is fit to a set of site arrival times,  and instead working directly with the profile data, these effects can be modelled appropriately, and the covariance with a possible background, or other signals of interest, can be significantly decreased.  As upcoming radio telescopes such as the Square Kilometre Array come on-line, improvements in sensitivity  will increase the significance of these effects on the pulse arrival times as measured through traditional methods, making a profile domain analysis the natural choice in order to improve the prospects for the detection of gravitational waves with pulsars.

Finally, the methodology we have described can be easily extended to account for shape variations associated with geodetic precession in relativistic  systems, such as the Hulse-Taylor Binary and the double pulsar system, in which pulsars show dramatic secular shape variability.   While these shape variations are not stochastic, by accounting for shape variability simultaneously to timing, a robust characterization of the relativistic orbit can be obtained, and the precision of tests of general relativity will be increased.

\section{Acknowledgements}

The Parkes radio telescope is part of the Australia Telescope National Facility which is funded by the Commonwealth of Australia for operation as a National Facility managed by Commonwealth Science and Industrial Research Organization (CSIRO).  We thank all of the observers, engineers, and Parkes observatory staff members who have assisted with the observations reported in this paper. 

LL was supported by a Junior Research Fellowship at Trinity Hall College, Cambridge University.  
RMS acknowledges travel support from the CSIRO through a John Philip award for excellence in early career research.

\bibliographystyle{mn2e}
\bibliography{references}

\end{document}